\documentclass[aps,prd,preprint,groupedaddress]{revtex4}
\usepackage{latexsym,amsfonts}
\usepackage{hyperref}
\usepackage{graphicx}

\begin{document}     

\title{The theory of surface heat capacity and its experimental verification. 
The scaling of specific heats of diamond lattice materials}

\author{Yuri Vladimirovich Gusev}

\affiliation{Department of Physics, Simon Fraser University, 8888 University Drive, Burnaby, B.C. V5A 1S6, Canada}
\affiliation{Lebedev Research Center in Physics, Russian Academy of Sciences, Leninsky Prospect 53, str. 11 (38), Moscow 119991, Russian Federation\\
Email: {\tt yuri.v.gussev@gmail.com}}

\date{\today}

\begin{abstract}
The  field (geometrical)  theory of specific heat is based on the universal thermal sum, a new mathematical tool derived from the evolution equation in the Euclidean four-dimensional spacetime, with the closed time coordinate. This theory made it possible to study the phenomena of scaling in the heat capacity of condensed matter. The scaling of specific heat of the carbon group elements with a diamond lattice is revisited. The predictions of the scaling characteristics for natural diamond and grey tin are verified with experimental data. The fourth power in temperature in the quasi-low temperature behaviour of the specific heat of both materials is confirmed. The phenomenon  of scaling in the specific heat of some zincblend lattice compounds and diamond lattice elements is explored with their characteristic temperatures. The nearly identical elastic properties of grey tin and indium antimonide is the cause for similarity of their thermal properties, which makes it possible to propose conjectures about the thermal properties of grey tin. The derivation of the specific heat of two-dimensional bodies is presented and used to explore the surface heat capacity. The surface specific heat, which is inversely proportional to the effective size of a material body, must always be considered in theory and experiment.  The surface contribution in total specific heat, which at sufficiently low temperatures is the cubic in temperature term, is demonstrated to be present in the datasets for powders of grey tin and sodium chloride, and two natural diamonds.  
\end{abstract}


\maketitle

\section{Thermal theory as geometry}
\label{theory}

\subsection{Overview of the problem}
\label{overview}

In the present paper we continue developing the {\em geometrical theory of specific heat} of condensed matter \cite{Gusev-FTSH-RJMP2016,Gusev-QLTBSH-RSOS2019} based on the finite temperature field theory \cite{Gusev-FTQFT-RJMP2015} and exploring the scaling phenomena found in experimental data for  materials with the diamond and zincblend lattices. This theory was called 'the field theory' in Refs. \cite{Gusev-FTSH-RJMP2016,Gusev-QLTBSH-RSOS2019}, because its mathematical formalism was used previously by quantum field theory. However, no  fields are considered here, the physical model deals with elastic (sound) waves in media. At its fundamental level, the model is build on the concept of geodesic distance, which belongs to geometry \cite{Synge-book1960}.  The theory takes as its input parameters geometrical characteristics of a matter system, the inter-atomic distance and the velocity of sound in condensed matter, and gives as the output an observable quantity, specific heat function.

The theory of specific heat of solid bodies was derived by P.J.W. Debye more than a century ago \cite{Debye-AdP1912}, at the time of the advent of quantum theory \cite{Poincare-JdP1912}, but before quantum field theory was created. The concept of four-dimensional space-time was already formulated \cite{Minkowski-PZ1908}, but these mathematical ideas were not yet widely used in physics. Yet, the Debye theory (or rather its elements) remains a standard model of condensed matter physics \cite{Kittel-book2005,Tsuji-book2010,Dekker-book1960}. The Debye theory is usually presented as a model for the {\em lattice} specific heat. In reality, this model was derived for and is applicable to {\em any elastic media}: crystalline, amorphous and liquid matter. 

However, the atomistic notion of discreteness of media was not properly built into the Debye model as explained in \cite{Gusev-FTSH-RJMP2016,Gusev-QLTBSH-RSOS2019}. The main physical idea underlying the Debye theory, namely, the correspondence between the standing elastic (sound) waves in a medium and a medium's heat capacity, could not be mathematically implemented at that time due to the lack of required mathematics. This early ideas of spectral geometry were  created by P. Debye and H. Weyl \cite{Weyl-RCMP1915}, but the evolution equation and its kernel were developed only in the second half of the 20th century \cite{CPT2}. P. Debye was forced to make his model closed by postulating its equivalence to the discrete model of Einstein's quantum oscillators \cite{Einstein-AdP1907}. This postulate cannot be verified experimentally and it is  incompatible with an elastic medium model. This and  other drawbacks of the Debye theory  rendered it practically useless. It fails to correctly describe matter's specific heat in any temperature region, so, experimental physicists and engineers rely on  data tables and  fitting equations, e.g. \cite{Mozgovoy-book2006}, instead of equations.

Theoretical physics must be {\em relativistic} \cite{Poincare-RCMP1906}. Therefore, any physical theory must be built in the four-dimensional {\em space-time}, introduced by H. Minkowski \cite{Minkowski-PZ1908}. The case of  electrodynamics was obvious and elaborated immediately by Lorentz, Poincar\'{e} and Einstein, but the application of this physical concept to thermal phenomena was delayed until the 21st century. The Planck constant was originally introduced by M. Planck in the context of thermodynamics, \cite{Planck-AdP1900},\cite{Planck-book1914},\cite{Planck-book1972}, p. 55, in order to find the empirical law of thermal radiation later named after him. Without one of four {\em defining} (formerly called 'fundamental') physical constants \cite{CODATA-Metro2018} it is impossible to construct consistent physical theories. In the New SI (2019) of physical units, the Planck constant is fixed as the defining constant of the unit of mass \cite{Stock-Metro2019}. We show how it can be used to connect mechanical (elastic) properties of matter with thermal ones \cite{Gusev-FTSH-RJMP2016,Gusev-QLTBSH-RSOS2019}.

\subsection{Geometrical formalism for thermal theory}
\label{fieldtheory}

We adopted the Debye's idea about heat as the energy density of standing acoustic waves in elastic bodies and proposed its {\em pseudo-relativistic} implementation in the four-dimensional space-time \cite{Gusev-FTSH-RJMP2016}. This physical model uses the geometrical formalism for finite temperature field theory \cite{Gusev-FTQFT-RJMP2015}. The theory  is based on new principles: 1) generating functionals of a physical theory are dimensionless and depend on dimensionless variables, 2) these functionals are related to physical observables  by the dimensional calibrating parameter determined by experiment. 

The mathematics of finite temperature field theory \cite{Gusev-FTQFT-RJMP2015} begins with the kernel of the evolution equation \cite{CPT2,Gusev-NPB2009},  
\begin{equation}
	\Big(\frac{\mathrm d}{{\mathrm d} s} - \Box^x \Big) K (s| x,x') = 0,    
\label{eveq}
\end{equation} 
with the intial conditions,
\begin{equation}
K (s| x,x') = \delta(x,x'), \ s/\sigma(x,x') \rightarrow 0. 
\label{Kxx}
\end{equation} 
(In quantum field theory Eq. (\ref{eveq}) used to be  called the 'heat equation', even though it is a fundamentally different equation.) The parameter of proper time $s$ came from geometrical analysis (differential geometry), and the world function $\sigma(x,x')$ is a square of the geodesic distance in the spacetime of dimension $D$ \cite{Synge-book1960,Gusev-FTQFT-RJMP2015}. The spacetime dimension $D$ is  embedded in the definition of the Laplacian $\Box$. The total spacetime of the dimension, $D=d+1$, is split into a product, $\mathbb{R}^d \times \mathbb{S}^1$, the ordinary space with the Euclidean time coordinate as a circle $\mathbb{S}^1$, with circumference $1/\beta$.

The fundamental solution for the evolution kernel,
\begin{equation}
K (s| x,x') = \frac{1}{(4 \pi s)^{D/2}} \exp\Big(-\frac{\sigma(x,x')}{2s} \Big),
\label{fundamentalK}
\end{equation}
is a  geometrical object. We need  the functional trace of the evolution kernel which is defined as the integral over the spacetime domain and the matrix trace operation (which is absent here but will be needed for future work, 
\begin{equation} 
	{\mathrm{Tr}} K(s) 
	\equiv 
	\int {\mathrm d}^{D} x \,   
	K(s|x,x) 
\label{TrK} 
\end{equation}
Finite temperature field theory \cite{Gusev-FTQFT-RJMP2015} is fundamentally based on the geometry of {$d$}-dimensional {\em spatial} domain of the $D$-dimensional spacetime. Its non-trivial topology is responsible for the introduction of temperature as the inverse of the circumference of the closed Euclidean time $\beta$,
\begin{equation}
\beta =  \frac{h}{k_{\mathsf{B}}}  \frac{v}{T}.
\label{beta}
\end{equation} 
Note, that this equation defines temperature, $T$, as the inverse of the Euclidean time length, $\beta$, with the dimensionality of length (m), not vice versa. In other words, geometry is fundamental and it determines thermodynamics (and temperature).

Therefore, the thermal sum is calculated in \cite{Gusev-FTQFT-RJMP2015},
\begin{equation} 
	{\textrm{Tr}}{K}^{\beta}(s)=  
	\frac{\beta}{(4 \pi s)^{1/2}}\
	\sum_{n=1}^{\infty} {\mathrm{e}}^{-\frac{\beta^2 n^2}{4s}}  \
	{\mathrm{Tr}} K(s).  \label{TrKb}
\end{equation}
The  trace of the $d$-dimensional evolution kernel ${\mathrm{Tr}} K(s)$ explicitly depends of the spacetime dimension (\ref{TrK}), e.g. in three dimensions it is,
\begin{equation}
	 {\mathrm{Tr}} K(s) = 
	\frac1{(4\pi s)^{3/2}}\ 
	\mathcal{V}, 
\label{TrKD3} 
\end{equation}
where $\mathcal{V}$ is the volume of the domain. 

The proper time integral that defines axiomatically the universal thermal sum (this integral was inspired, but not derive, by the finite temperature quantum field theory) is,
\begin{equation}
-F^{\beta}_{\tilde{a}} \equiv {\tilde{A}}
 \int_{\tilde{a}^2/4}^{\infty}\! \frac{{\mathrm d} s}{s}\,  
	{\textrm{Tr}}  K^{\beta}(s).         
\label{Fbeta}
\end{equation} 
The crucial difference here is that no quantum fields are considered here even implicitly, and the lower limit of the integral comes from the physical restriction on the existence of short-length (high-frequency) sound waves in atomistic condensed matter, which is not an ideal elastic medium.  

The computation of (\ref{Fbeta}) delivers \cite{Gusev-FTSH-RJMP2016}, up to the overall  factor, the dimensionless expression,
\begin{equation}	   
	-F (\alpha) =
	\frac{\tilde{A}}{\pi^2}
	\frac{\mathcal{V}}{\tilde{a}^3}
	\sum_{n=1}^{\infty}\, \frac{1}{n^4 \alpha^3}\Big(1 -\exp (-\alpha^2 n^2) 
	- n^2 \alpha^2 \exp (-\alpha^2 n^2)\Big), \ (d=3),
	\label{TF}
\end{equation}
of the dimensionless variable, 
\begin{equation}
\alpha =  \frac{1}{B} \frac{h}{k_{\mathsf{B}}}  \frac{v}{a T}.
\label{alpha}
\end{equation}
Here the Planck constant $h= 6.626 070 15 \cdot 10^{-34}\ J\cdot s$ and the Boltzmann constant $k_{\mathsf{B}}=1.380\, 649 \times 10^{-23}\ J K^{-1}$ are exact by the New SI, while $a$ is the average inter-atomic distance (or lattice constant), $v$ is the group velocity of sound, and $B$ is the experimental calibration parameter. This is it, the theory, with a single velocity of sound, is calibrated by two parameters, $A$ and $B$, which scale the sought function horizontally and vertically.

Ref. \cite{Gusev-FTSH-RJMP2016} proposed a conjecture that  the universal thermal sum $F (\alpha)$ to be the chief functional of new, geometrical, thermal theory. This declaration resembles thermodynamic ensembles of the theory of J.W. Gibbs \cite{Gibbs-book1902}. Traditional thermodynamics is an  axiomatic   theory and its potentials are not directly connected to physical observables. Presented  is a different axiomatic theory, which is based on a different branch of mathematics. Despite some similarity, this sum  is quite different from the sums of traditional thermodynamics. On one hand, no notion of energy is used here, on the other hand, the $\alpha^2$ in the exponents of $F (\alpha)$ is proportional to a square of the sound velocity, thus, it could be associated with the kinetic energy of elastic waves in matter. We eliminated the physical time in theory, since temperature, $T$, is expressed now with the inverse of the closed Euclidean time, $\beta$. This is a natural step, because we want to describe phenomenology of what is usually called 'thermal systems in equilibrium'.  This step should be modified if we want to describe time-dependent phenomena, like non-stationary heat conductivity. At the same time, physical time is implicitly present in the theory through the velocity of sound, $v$, in its variable (\ref{alpha}).

The functional $F (\alpha)$, and its derivative, are dimensionless because both, the integrand (\ref{TrK}) and the measure of the proper time integral  (\ref{Fbeta}), are dimensionless. This fact poses the fundamental problem of finding a way to obtain physical (dimensional) quantities, called observables, from a mathematical (pure number) expression (\ref{Fbeta}). We used the only feasible way and introduced a dimensional factor by the scaling postulate \cite{Gusev-FTSH-RJMP2016}. This postulate might look odd, but in fact it is the most natural and common way of producing physical equations. Indeed, any  fundamental equation in theoretical physics, from the Newton's law of gravitation to the Dirac equation, contains dimensional parameters, which are called physical constants. They match the mathematical structure of a physical theory to experimental measurements. Physical {\em models}, which are limited in the scope of their applications, contain many more special physical constants that are called calibration parameters.

Due to the scaling postulate, the volume specific heat function is obtained by the derivative of (\ref{TF}) over its variable $\alpha$,
\begin{equation}
C_{\mathcal{V}} \equiv k_{\mathsf{B}} \frac{\partial F(\alpha)_{3d}}{\partial \alpha},   
\label{CV3d}
\end{equation} 
or the molar specific heat (at constant pressure) as derived in \cite{Gusev-FTSH-RJMP2016},
\begin{equation}
	C = A
	k_{\mathsf{B}} N_{\mathsf{A}}\, 
	\Theta(\alpha), 
	\label{CMA}
\end{equation}
where the Avogadro constant $N_{\mathsf{A}}=6.022\, 140\, 76 \times 10^{23}\ {mol}^{-1}$  gives with $k_{\mathsf{B}}$ the usual gas constant, $R = 8.314\, 734\ J\, mol^{-1}\, K^{-1}$.
The dimensionless thermal function in (\ref{CMA}) is calculated from (TF) as,
\begin{equation}
\Theta(\alpha) =
\sum_{n=1}^{\infty}\,
	\frac{1}{ n^4 \alpha^4} 
	\Big\{
	 1 -\exp (-\alpha^2 n^2)
	 - n^2 \alpha^2\,  \exp (-\alpha^2 n^2)
	- \frac{2}{3} n^4 \alpha^4 \,  \exp (-\alpha^2 n^2)
	\Big\}, \ d=3.
	\label{Theta}
\end{equation}

\begin{figure}[ht]
  \caption{$\Theta (\alpha)$ vs $1/\alpha$, $d=3$}
  \centering
\includegraphics[scale=0.5]{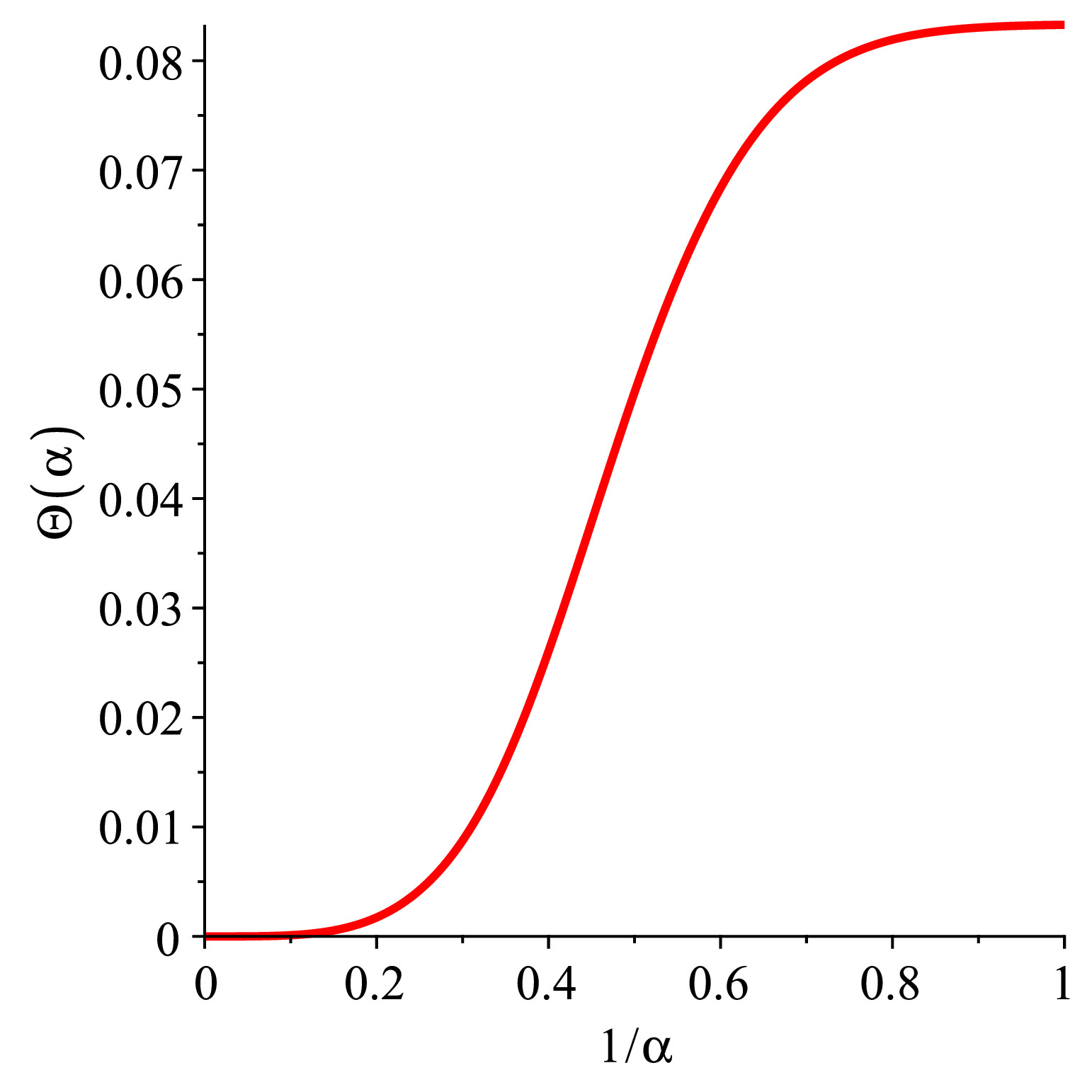}
\label{Theta-alpha-3d}
\end{figure}

The molar heat capacity (\ref{CMA}) is defined and measured at constant pressure, because the inter-atomic distance and the sound velocity can change with changing pressure, but the theory's input parameters are assumed to be constant the variable $\alpha$. This is a contribution from a single velocity $v_i$ supported within elastic medium: the theory is not completed yet and it suffers the same shortcoming as the original Debye theory. Indeed, this axiomatically derived function possesses the behaviour which is too good not to use in the thermal theory. In Fig. \ref{Theta-alpha-3d}
 $\Theta(\alpha)$ is plotted as a function of the inverse variable, $1/\alpha$, which mimics the temperature due the definition (\ref{alpha}). The assumption that $v$ and $a$ remain approximately constant with the temperature change is satisfied in the first approximation, as  discussed in \cite{Gusev-QLTBSH-RSOS2019}. This condition maybe violated in the pre-melting region due to thermal expansion and changing elastic moduli, but we do not consider this limit here. 
 
At sufficiently low temperatures, function  (\ref{Theta}) simplifies to power-like contributions. In this limit, which we called the quasi-low temperature regime \cite{Gusev-QLTBSH-RSOS2019}, only the slowest  velocity of sound dominates in the total specific heat; for diamond lattice materials this velocity is $v_5$ determined by $c_{44}$ elastic modulus. The quasi-low temperature behaviour of specific heat \cite{Gusev-QLTBSH-RSOS2019} is the asymptotics, $\alpha \rightarrow \infty$, of the function (\ref{CMA}),
\begin{equation}	   
	C \propto k_{\mathsf{B}} N_{\mathsf{A}}
 \frac{1}{\alpha^4}, \ \alpha \rightarrow  \infty . 
 \label{CLTi}
\end{equation}
As seen  from the definition (\ref{alpha}), this behaviour predicts the fourth power of temperature \cite{Gusev-FTSH-RJMP2016}.  This quartic in temperature contribution (\ref{CLTi}) in the total specific heat was found in some experimental data sets \cite{Gusev-FTSH-RJMP2016,Gusev-QLTBSH-RSOS2019}. In the present work, we verify some conclusions made in the first paper \cite{Gusev-FTSH-RJMP2016} for the diamond lattice materials and  supplement it with the analysis of some zincblend compounds. 

From the dimensionless (scaling) property of the universal thermal function (\ref{Theta}) of the dimensionless variable $\alpha$ we concluded \cite{Gusev-FTSH-RJMP2016} that the threshold for the change in the functional behaviour can be identified from the characteristic dimensionless value,
\begin{equation}	   
	\alpha_{0} = \frac{h}{k_{\mathsf{B}}}  \frac{v}{a T_0}. 
\label{alpha0}
\end{equation}
which should be the same for the class of materials with the same kind of lattice. 
Note, that the notation $\alpha_0$ was introduced in \cite{Gusev-QLTBSH-RSOS2019} and replaced the  notation $\theta$ of Ref. \cite{Gusev-FTSH-RJMP2016} in order to avoid confusion with the $\Theta$ functional and to reflect its proper meaning.

Different values of the characteristic temperature $T_0$ can be then calculated with the known atomic and mechanical properties of a material and used to {\em empirically calibrate} specific heat functions. In other words, the {\em hypothesis} is that specific heat of only one material should be measured, while other functions of the same lattice class could be described by the same mathematical function, whose actual  values can be found once $T_0$ for a material not measured is calculated from (\ref{alpha0}). The value of the specific heat at the characteristic temperature, $C_0$, is expected to be the same for all materials in the same lattice class. This procedure, when the theory is completed and calibrated, could greatly reduced the amount of experimental work needed. In Ref. \cite{Gusev-QLTBSH-RSOS2019} we  explained that the characteristic (\ref{alpha0}) is similar to or even coincide with some other characteristics introduced in different methodological approaches.

The Debye temperature, which originally used to be considered as the universal characteristic of a specific kind of matter, became really a different form for presenting measured specific heat. Indeed, it is very convenient for the analysis of the power behaviour of molar heat capacities. To indicate the fact that is merely a testing function of experimental data, we removed all unessential numerical factors and defined it as \cite{Gusev-FTSH-RJMP2016},
\begin{equation}	   
	T_{\theta} \equiv T \left( \frac{R}{C}\right)^{1/3}. 
\label{thetaT}
\end{equation}
It can be used to find the characteristic temperature, $T_0$, in the same way the standard graph of $C/T^3$ vs $T$ is used. But the latter graph is more convenient for finding  numerical values.

\subsection{Surface specific heat from the field theory in the 2+1 dimensional spacetime}
\label{3Dtheory}

The hypothesis that {\em the surface contribution to specific heat} of solid bodies should obey, in the quasi-low temperature regime, the cubic in temperature law is one of the main predictions of the field theory of specific heat \cite{Gusev-FTSH-RJMP2016}. In this theory, the quasi-low temperature contributions from the hierarchy of geometrical characteristics of a three-dimensional body (volume, surface, and edges) form the decreasing powers of temperature: $T^4$, $T^3$, $T^2$. Therefore, one should expect to observe in the specific heat of a solid body with {\em non-smooth} surface the second order of temperature, beside the quartic and cubic ones. However, both, cubic and quadratic in temperature terms in the specific heat are not universal and depend and the size and shape of a body.

The universal thermal function for two-dimensional manifolds, which can be also a submanifold of the three-dimensional spacetime a surface, is easily calculated using the formalism of the previous section, which dealt with three spatial dimensions. We use the equation defining the universal thermal sum as the integral over proper time with the lower bound (\ref{Fbeta}) of the finite temperature evolution kernel trace (\ref{TrKb}). These equations are  the same, but the trace of the evolution kernel  is different,
\begin{equation} 
	{\mathrm{Tr}} K(s) 
	 =  \frac1{(4\pi s)}\
 	\mathcal{S},  
 	\label{TrKD2} 
\end{equation}
where $S$ is the two-dimensional volume of a spatial domain. This could be an area of a surface of a three-dimensional condensed matter body. This is it, we assume that the functional of heat capacity of a boundary is independent of the bulk heat. Of course, it is an approximation which should be improved because elastic waves inside a body and on its surfaces are coupled. However, the calculations below give an exact result, if one deals with (suspended) two-dimensional materials, the popular subject nowadays. We note that the presented behaviour is universal to any sound velocity, including the out-of-plane waves, but the lowest velocity of sound always dominates the sum.

With the variable change $y=\beta^2/(4s)$, the defining integral (\ref{Fbeta}) turns out to be,
\begin{equation}
-F(\alpha) _{2d} = 
	\frac{\tilde{D}}{\pi^{3/2}}
	\frac{\mathcal{S}}{\tilde{a}^2} 
		\sum_{n=1}^{\infty}\, 
		\int_{0}^{\alpha^2} \mathrm{d} y \ y^{1/2} \mathrm{e}^{-y n^2}.    
\label{Fbntegral2d}
\end{equation} 
(The $d=3$ case \cite{Gusev-FTSH-RJMP2016} of this equation had $y^1$ in the integral.)
The ratio $\frac{\mathcal{S}}{\tilde{a}^2} $ is proportional to the total number of atoms on a surface.
The result of the integration in (\ref{Fbntegral2d}) is,
\begin{equation}
-F(\alpha) _{2d} = 
	\frac{\tilde{D}}{\pi^{3/2}}
	\frac{\mathcal{S}}{\tilde{a}^2} 
		\sum_{n=1}^{\infty}\, 
		\frac{1}{n^3 \alpha^2}
		\Big\{
	\int_0^{\alpha n} \mathrm{d} t \ \mathrm{e}^{-t^2} 
	 - n \alpha\,  \mathrm{e}^{-\alpha^2 n^2}
	\Big\}        
\label{Fantegral2d}
\end{equation}

\begin{figure}[ht]
  \caption{$\Theta (\alpha)$ with $n=1$ vs $\alpha$, $d=2$}
  \centering
\includegraphics[scale=0.5]{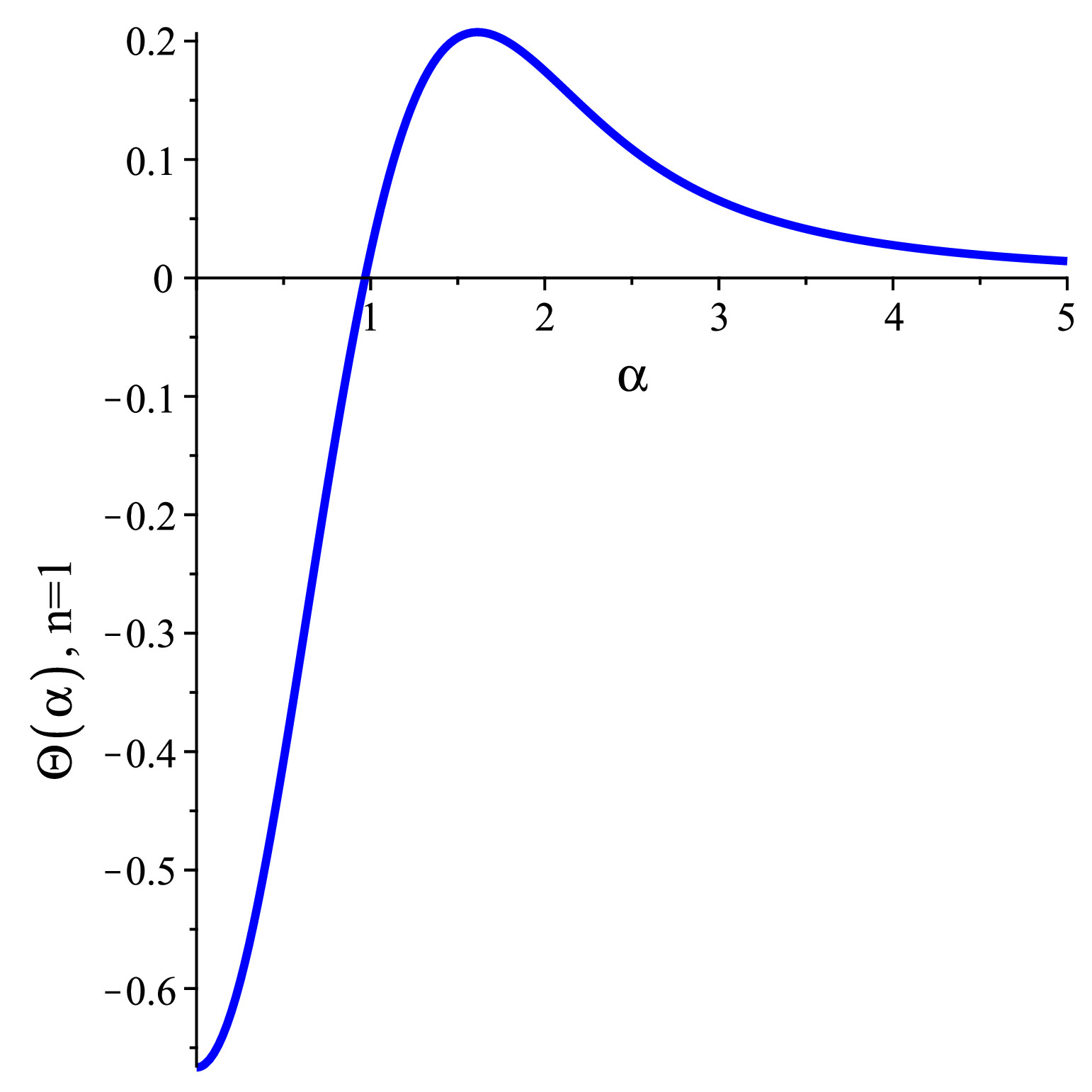}
\label{Theta-alpha-N1-2d}
\end{figure}

We take the derivative of (\ref{Fantegral2d}) over the intrinsic variable $\alpha$ and obtain an expression where the sum above becomes,
\begin{equation}
\Theta(\alpha)_{\mathrm 2d} =
 \sum_{n=1}^{\infty}\,
	\frac{1}{ n^3 \alpha^3} 
	\Big\{
	\sqrt{\pi} \ \mathrm{erf}(\alpha n)
	 - 2 n \alpha\,  \exp (-\alpha^2 n^2)
	- 2 (n \alpha)^3 \,  \exp (-\alpha^2 n^2)
	\Big\},
	\label{Theta2D}
\end{equation}
where the standard definition for the error function,
\begin{equation}
\sqrt{\pi}\ \mathrm{erf}(\alpha n) \equiv  2 \int_0^{\alpha n} \mathrm{d} t \exp(-t^2) .
\label{errorf}
\end{equation}
The basic function of the sum (\ref{Theta2D}) ($n=1$) is displayed in Fig. \ref{Theta-alpha-N1-2d}.
\begin{figure}[ht]
  \caption{$\Theta (\alpha)$ with $n=1000$ vs $\alpha$, $d=2$}
  \centering
\includegraphics[scale=0.5]{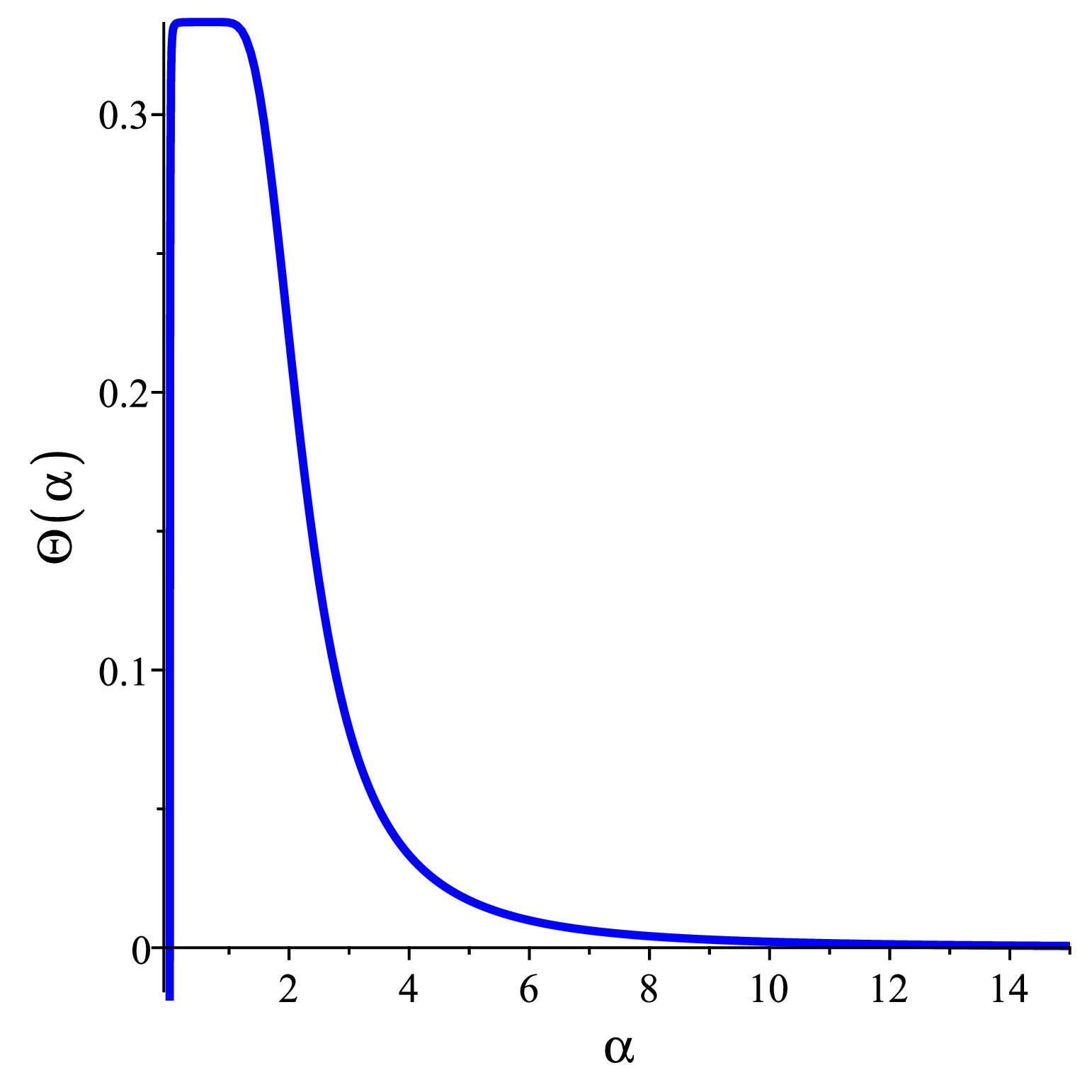}
\label{Theta-alpha-N1000-2d}
\end{figure}
The sum converges quite quickly and for $n=1000$ is evaluated in Fig. \ref{Theta-alpha-N1000-2d}. The plot's horizontal axis begins at $\alpha=0.05$ because the thermal sum diverges as $\alpha \rightarrow 0$. This behaviour is expected since the variable $\alpha$ cannot accept zero value due its definition. Physically this means that the wave approximation breaks down at certain high frequencies due to the  atomic discreteness of condensed matter. This breakdown is a benign feature of the theory because condensed matter should undergo a phase transition at some high temperature.

\begin{figure}[ht]
  \caption{$\Theta (\alpha)$ vs $1/\alpha$, $d=2$}
  \centering
\includegraphics[scale=0.5]{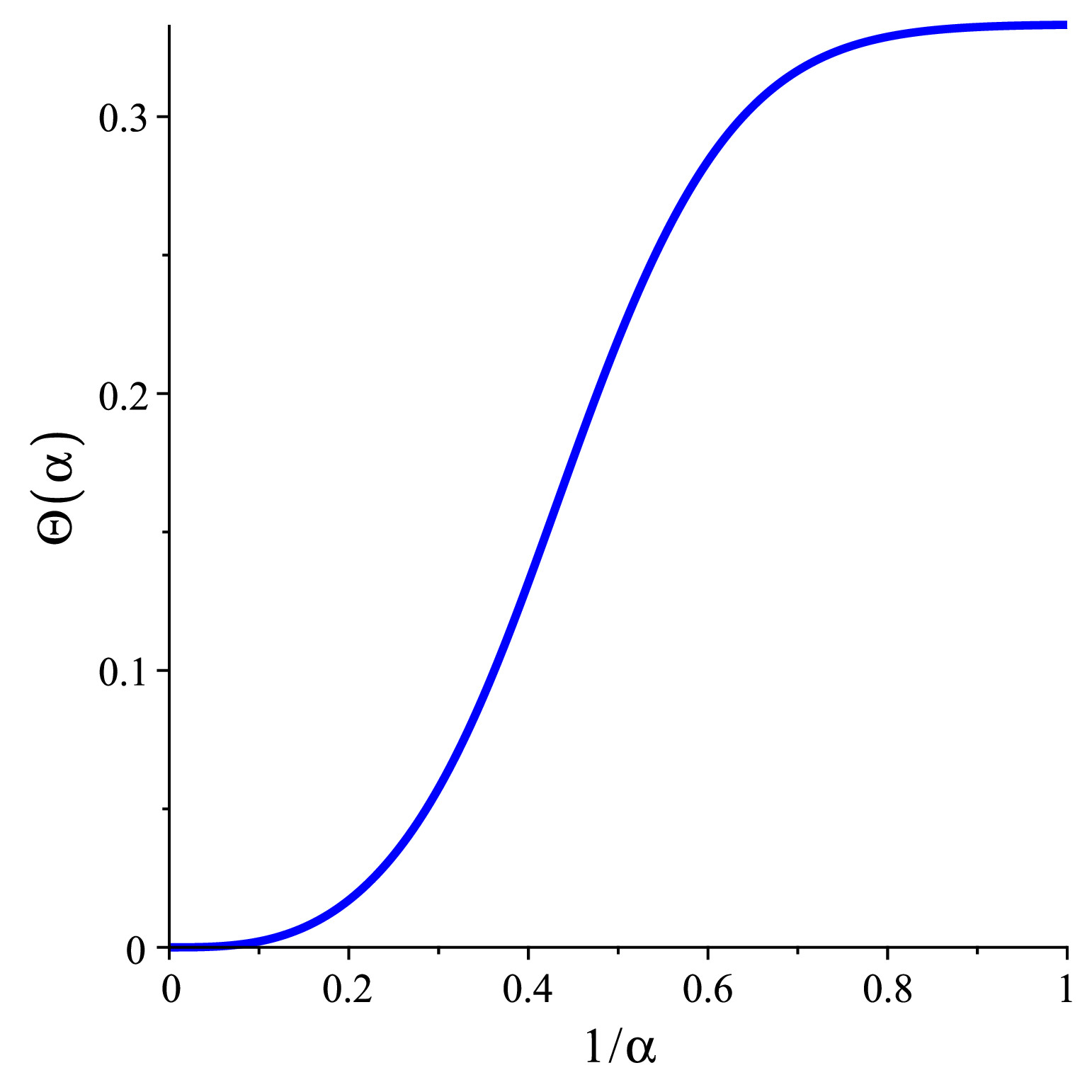}
\label{Theta-1alpha-2d}
\end{figure}

If we look at the universal thermal function in the inverse variable $1/\alpha$, it accepts, in the region of our interests, a familiar form, Fig. \ref{Theta-1alpha-2d}. Note, that this graph is qualitatively similar to Fig. \ref{Theta-alpha-3d}, but its function is quite different (\ref{Theta2D}). At higher values of its argument, this graph would decrease and go to negative values, as determined by the $\alpha \rightarrow 0$ breakdown in Fig. \ref{Theta-alpha-N1000-2d}, but this is irrelevant to physics. 
The scaling postulate of \cite{Gusev-FTSH-RJMP2016} is expressed in two-dimensional case as the specific heat for a fixed area, $S$, as,
\begin{equation}
C_{\mathcal{S}} \equiv k_{\mathsf{B}} \frac{\partial F(\alpha) _{2d}}{\partial \alpha}.   
\label{CS2d}
\end{equation} 
The expression (\ref{Fantegral2d}) contains the ratio, which can be interpreted as the total number of atoms of the two-dimensional matter, $\mathcal{S}/\tilde{a}^2 \propto N = n N_{\mathsf{A}}$. Indeed, the cut-off parameter is proportional to the characteristic length scale, which is the average inter-atomic distance, $a$. The initially used arguments about the lattice constant and the number of nodes in a unit, which were used to derive the specific heat for the cubic lattice \cite{Gusev-FTSH-RJMP2016} are not restrictive, because there is an overall calibration factor in (\ref{Fantegral2d}) and (\ref{CS2d}). Therefore, we keep only one undetermined parameter in the molar specific heat after dividing (\ref{CS2d}) by the number of moles $n$,
\begin{equation}
C_{\mathrm{2d}} = D k_{\mathsf{B}} N_{\mathsf{A}} \Theta(\alpha)_{\mathrm{2d}}.   
\label{C2d}
\end{equation}

The leading contribution in the specific heat of two-dimensional (graphene etc) materials should exhibit, in the quasi-low temperature regime, the {\em cubic} in temperature behaviour, 
\begin{equation}
C_{\mathrm{2d}} \propto k_{\mathsf{B}} N_{\mathsf{A}} T^3, \ T < T_0\ (d=2), 
\label{2D}
\end{equation}
while its sub-leading asymptotics is the quadratic in temperature. The asymptotic behaviour (\ref{2D}) contradicts to the result derived within the Debye theory for lower dimensional systems \cite{Gurney-PR1952}, where the quadratic law for the intrinsic ('bulk', i.e. independent of the shape and size of flakes) specific heat of two-dimensional systems is proposed, as accepted by some modern researchers \cite{Pop-MRSB2012}.

Let us also expose the history of this subject. Apparently, Tarasov was first to propose that one-dimensional systems (very long molecules within polymers) and two-dimensional-systems (single and multi-layer flakes of graphene within graphite) are described by the low temperature Debye laws corresponding to the linear order in $T$ for the former case \cite{Tarasov-DAN1945-1} and the second order in $T$ for the latter case \cite{Tarasov-DAN1945-2}. The English translation of Ref. \cite{Tarasov-DAN1945-2} is cited in the literature and its derivation is reproduced in the textbook of Semenchenko \cite{Semenchenko-book1966}, Eq. 73.15 on p. 324. Even though results of these works are {\em wrong} by one order of temperature, because the Debye theory itself is wrong as explained in \cite{Gusev-FTSH-RJMP2016,Gusev-QLTBSH-RSOS2019}, we refer to them as an early example of the geometrical hierarchy of lower dimensional spaces studied in the present work. 

Despite the large number of experimental studies of graphene, to our knowledge, no direct measurement of the specific heat of suspended graphene has been done yet. Thus, conflicting hypotheses could be verified by experiment. Perhaps, indirect experimental conclusions can be extracted from the measurements of graphite's specific heat as discussed in the literature. However, supposed confirmations are rather vague, because they are often made with the use of the logarithmic scale plots,  which are known to confuse statistical conclusions. For example, see work \cite{Broido-NC2019}, which dispelled the illusion of the scale-free networks in nature and society that persisted for about two decades, mainly due to the log-log plot analysis.

We add that the sub-leading power, determined by the edges of graphene's flakes  indeed obeys the quadratic law at even lower temperature, but this contribution is {\em not} universal. Because such a quadratic contribution, $T^2$, depends on the size and the shape of graphene's flakes, it would have different absolute magnitudes (coefficients at the $T^2$ term), which is apparently demonstrated by the existing experimental data for graphite.

Let us now consider the surface contribution to the total specific heat of a three-dimensional body, under an assumption that a surface one-atomic layer can be described as a two-dimensional body, which gives an additive term. First, we introduce the {\em effective size} of a body, which is defined in \cite{Gusev-ILTR-RJMP2014} as the ratio of the surface area $\mathcal{S}$ to the volume $\mathcal{V}$,
\begin{equation}
r \equiv \mathcal{V}/\mathcal{S}.
\label{effsize}
\end{equation}
Then, the total  heat capacity is defined as a sum of two independent contributions,
\begin{equation} 
	C_{\mathrm{tot}} 	 = C_{\mathcal{V}} + C_{\mathcal{S}},  
	\label{Ctotal} 
\end{equation}
There is no a priori principle that would fix the relative constant of these two terms, which could be another calibration parameter. However, since both terms have their own calibration parameters, $A$ and $D$, the third one is redundant; we use below the set of $A$ and $C$, the relative coefficient. The surplus of the surface heat is treated as a component of the volume (bulk) heat, therefore, the surface specific heat is divided by the number of moles, $n_{\mathcal{V}}$, of the whole sample with volume $\mathcal{V}$. As a result, the surface term is proportional to the inverse effective size of a body,
\begin{equation} 
	C_{\mathrm{tot}} 	 =  A k_{\mathsf{B}} N_{\mathsf{A}} 
	\Big( \Theta(\alpha)_{\mathrm 3d} + C r^{-1}  \Theta(\alpha)_{\mathrm 2d} \Big). \label{Ctotalr} 
\end{equation}
We apply here a crude approximation that the transverse velocities of sound in bulk and on surface are equal (comparable). Only  then we can assume the same variable $\alpha$ and take the derivative of both, volume and surface, term over it. Reference tables confirm this approximation is acceptable and it can be elaborated in future. 

Using the quasi-low temperature asymptotics obtained above we conjecture that the quasi-low temperature behaviour of specific heat of a three-dimensional body of volume $\mathcal{V}$ with a smooth surface of an area $\mathcal{S}$ is,
\begin{equation} 
	C_{\mathrm{tot}} 	 =  A_{\mathrm{QLT}} k_{\mathsf{B}} N_{\mathsf{A}} 
	\Big( T^4 + C_{\mathrm{QLT}} r^{-1} T^3 \Big), 
	T < T_0.	
	 \label{CtotalLT} 
\end{equation}
where we omit unessential numerical coefficients to elucidate the power like terms. The formula (\ref{CtotalLT}) is tested in Sect. \ref{surface} with some specific heat datasets.

\section{Experimental verification and scaling} 
\label{verification}

The characteristic temperature $T_0$ and the parameter $\alpha_0$ for natural diamond, derived from the experimental data obtained by J.A. Morrison's group \cite{Desnoyers-PhilMag1958}, were different from the ones for other carbon group elements \cite{Gusev-FTSH-RJMP2016}. The power law of diamond's specific heat in the quasi-low temperature regime was not determined. It was conjectured that  higher precision measurements of natural diamond's properties could produce new values that would be in agreement with the characteristics of silicon and germanium. The same characteristics for grey  tin, {$\alpha$-Sn}, were theoretically derived from its elastic properties, rather than determined from the specific heat data, thus  they should be verified. The quasi-low temperature behaviour of the diamond lattice and zincblend materials should be quantitatively similar, as was shown with GaAs. The experimental verification of these predictions is the subject of the next section.

\subsection{Physical properties of the diamond type lattice materials}
\label{scaling}

Thermal properties of the diamond lattices supported the idea of scaling, which axiomatically emerged from the field theory of specific heat \cite{Gusev-FTSH-RJMP2016}. The studied materials were three elements of the carbon group, diamond (C), silicon ($\alpha$-Si) and germanium ($\alpha$-Ge), and a chemical compound, gallium arsenide (GaAs).  We reproduce here from \cite{Gusev-FTSH-RJMP2016} the table of the physical properties of these materials, Table \ref{group4}, with the addition of other compounds with the zincblend lattice: indium antimonide (InSb), gallium antimonide (GaSb), and indium arsenide (InAs).

In Table \ref{group4} the parameters predicted in \cite{Gusev-FTSH-RJMP2016}  are replaced by measured values, while parameters that are not directly measured are marked in italic. We added the columns for the specific heat, $C_0$, at the characteristic temperature $T_0$ and the scaled coefficient $\tilde{d}_1$ of the fitting equation (\ref{linear}), as discussed in Sect. \ref{scaling}.
\begin{table}[!ht]
\caption{Physical properties of the carbon group elements and the zinc-blend compounds}
\label{group4}
\begin{tabular}{lllllllllllll}
\hline
material& a & $c_{11}$&$c_{12}$& $c_{44}$&$\rho$& $v_5$ & $T_0$&  $\alpha_0$ & $C_0$ & $\tilde{d}_1$ &$T_{\mathrm{m}}$& $C_{\mathrm{DP}}$\\
\hline
units& \AA & GPa & GPa & Gpa & g/cm${}^3$ & km/s & K & -- & J/(K mol) & J/(K mol) & K & J/(K mol)\\
\hline
diamond  & 3.567 & 1080.8 & 125.0 & 578.9 & 3.512 & 11.66 &{\em 161} & {\em 1.55}& {\em 1.25} & {\em 0.88} & -- & {\em 29.0} \\
$\alpha$-Si& 5.431 & 165.8 & 63.94 & 79.63 & 3.329 & 4.674 & 39.4 & 1.67& 1.189 & 1.42 & 1685 & 29.16 \\
$\alpha$-Ge& 5.657 & 128.5 & 48.26 & 66.80  & 5.3256 & 2.746 & 21.4 & 1.73 & 1.133 & 1.40 & 1210 & 28.76 \\
$\alpha$-Sn& 6.489 & 66.7 & 36.5 & 30.2 & 5.7710 & {\em 1.62} & 12.0 & 1.59 & {\em 1.176} & {\em 1.03} & {\em 800} & {\em 28.0} \\
InSb& 6.489 &  66.7 & 36.5 & 30.2 & 5.771 & 1.62 & 11.0 & 1.73 & 1.015 & 1.25 & 800 & 28.00 \\
GaSb& 6.096 &  88.3 & 40.2 & 43.2 & 5.614 & 2.07 & 15.0 & 1.73 & 1.033 & 1.24 & 985 &  28.38 \\
InAs& 6.058 &  83.4 & 45.4 & 39.5 & 5.68 & 1.83 & 14.0 & 1.65 & 1.047 & 1.25 & 1215 &  \\
GaAs& 5.653 & 118.8 & 53.7 & 59.4 & 5.32 & 2.47 & 21.0 & 1.67 & 1.187 & 1.25 & 1513 & 29.08\\
\hline
\end{tabular}
\end{table}

The value for the lattice constant of natural diamond is from \cite{Stoupin-PRL2010}, while diamond's elastic constants and  velocity are from \cite{McSkimin-JAP1972}. These measurements were later confirmed by \cite{Betts-JAP2008}. The values derived from the specific heat data of the zincblend compounds in Table \ref{group4} are from Ref. \cite{Cetas-PR1968}. The actual measurement of the elastic moduli of GaAs was made in work \cite{Bateman-JAP1959}. 

It is important to note that the molar specific heats of chemical compounds here and in \cite{Cetas-PR1968} are given {\em per number of atoms}, as fundamental constituents of condensed matter, not per number of molecule, i.e. one mole is equal to $N_{\mathsf{A}}$ of atoms even for a material with many-atomic molecules. Therefore, data taken from some other references should be divided by 2 to make them consistent with this rule.

As different from \cite{Gusev-FTSH-RJMP2016}, table \ref{group4} contains the melting temperature, $T_m$, not a general critical temperature, which previously included the ablation of diamond and the lattice transformation of grey tin. The melting of diamond occurs at temperatures approaching 5000 K, under high pressure \cite{Savvatimsky-book2014}, so we leave this cell blank. 

According to the field theory of specific heat, the {\em slowest} velocity of sound, which is one of transverse velocities, dominates the specific heat at low temperature \cite{Gusev-FTSH-RJMP2016}. For the diamond lattice, it is the transverse velocity $v_5$, in the notations of \cite{McSkimin-JAP1972}. The universal link between the transverse velocity of sound and the low temperature behaviour of specific heat of glasses was discovered long time ago, e.g. \cite{Shintani-NatMat2008}. This is an example of the same phenomena we continue to study  here with the diamond and zincblend lattice data.

In \cite{Gusev-QLTBSH-RSOS2019}, the linear  fit to the scaled specific heat function $C/T^3$ was introduced,
\begin{equation}
C/T^3 = d_0 + d_1 T, \ T<T_0. \label{linear}
\end{equation}
Obviously, this form implies the presence of a $T^3$ contribution in specific heat, which must exist due to the surface heat capacity, as hypothesised  in \cite{Gusev-FTSH-RJMP2016}. Regretfully, the topic of surface specific heat was excluded from the final version of Ref.  \cite{Gusev-QLTBSH-RSOS2019} (yet mentioned in the analysis of the fitting function  (\ref{linear}) on p. 10) on  the request from an anonymous referee (see the publication's history and its earlier versions in arxiv.org). This exclusion caused misunderstanding of the field theory of specific heat among condensed matter experts. This fact reflect the negative side of the journal peer reviewing process.

The consideration of a condensed matter system as a closed spatial domain with a smooth boundary, existing in the four-dimensional space-time with the cyclic Euclidean time, is a core mathematical concept of the finite temperature field theory \cite{Gusev-FTQFT-RJMP2015}. The boundary (a surface for a 3-d body or an edge for a 2-d sample) of a condensed matter system determines its physics. The surface specific heat is a dominant contribution at temperature lower than the quartic power (quasi-low temperature) regime. But even in the quasi-low temperature regime \cite{Gusev-QLTBSH-RSOS2019}, the absolute value of quartic function (\ref{CLTi}) may be comparable to other contributions, which are always present in experimental data, as is shown below. Furthermore, the full function of the surface specific heat is  the exponential sum (\ref{TF}), and we only consider the leading contribution of the $v_5$ velocity, while other modes do contribute as well.

The fact that there is always a cubic contribution present in the specific heat (\ref{linear})  means that first statistical estimates done in \cite{Gusev-FTSH-RJMP2016} were not complete. Is is clear that even if one were to expect the specific heat behave strictly as a $T^4$ function, this function should have the form $C=a+(T-b)^4$ because the origin of this quartic function is not at $T=0$, which is a forbidden temperature value in the finite temperature field theory \cite{Gusev-FTQFT-RJMP2015}. Then, this would be a full polynomial of the fourth order, with a constraint on its coefficients. Without the complete theory of specific heat yet, the following combination can be used as an approximation,
\begin{equation}
C = d_3 T^3 + d_4 T^4, \ T<T_0. \label{quartic}
\end{equation}
It is consistent with (\ref{linear}), but not equivalent to it. The coefficients of (\ref{linear}) and (\ref{quartic}) could coincide only within statistical uncertainty, $d_0 \approx d_3$ and $d_1 \approx d_4$, because the fitting is done respectively to scaled and original specific heat data.

For silicon and germanium, which were  considered in \cite{Gusev-FTSH-RJMP2016} (but only the germanium's analysis was reported), we used the data from work \cite{Flubacher-PM1959} and treated them with the above fitting equations. Silicon gives the coefficients $d_1 = 5.89(20) \cdot 10^{-7} \ J/(K^5 \cdot mol)$ and $d_0 = 2.3(4.2) \cdot 10^{-7} \ J/(K^4 \cdot mol)$. The standard errors are shown in the round brackets, i.e. $5.89(20) \equiv 5.89 \pm 0.20$. The very large standard error for $d_0$, whose range of acceptable values includes zero, means that the $T^3$ contribution is negligible. The coefficients for  germanium came up as $d_1 = 6.68(30) \cdot 10^{-6} \ J/(K^5 \cdot mol)$ and $d_0 = -6.64(4.25) \cdot 10^{-6} \ J/(K^4 \cdot mol)$, which is confronted with $d_4 = 6.01(37) \cdot 10^{-6} \ J/(K^5 \cdot mol)$ and $d_3 = 3.41(5.97) \cdot 10^{-6} \ J/(K^4 \cdot mol)$. Again, statistical significance of the coefficient of $T^3$ is not satisfactory due to the errors, and we put $d_3$ to zero.
This means, according to the field theory of specific heat, that experiments measured really bulk properties of the substances, and the surface heat was small. Indeed, the work \cite{Flubacher-PM1959} says that single crystal specimens were broken into pieces with the average size of 3 mm.

\subsection{Testing the theory with the grey tin data}  \label{tin}

Like carbon, silicon and germanium, tin is also an element of group IV of the Mendeleev's periodic table of chemical elements, whose 150th anniversary was celebrated last year.  Apart from allotropes created at high pressures, two forms of tin exist at pressures and temperatures available outside a lab. White tin ($\beta$-tin) is a metal with the body-centred tetragonal (bct) lattice structure. At temperature lower than 286.4 K it turns  into semiconductor, grey (gray) tin ($\alpha$-tin) with the cubic lattice of the diamond type \cite{Busch-SSP1960}. This transition is commonly  known as 'tin pest', a damaging factor in technological applications \cite{Burns-JFAP2009}. 

While thermal properties of  $\beta$-tin are well studied \cite{Abrikosov-Calphad2019}, $\alpha$-tin's specific heat remains poorly measured. Nevertheless, the acoustic frequencies of grey tin lattice were determined  in the neutron scattering experiments \cite{Price-PRB1971}.
Incidentally, the frequencies measurements \cite{Price-SSC1969} done with the neutron scattering were fitted with 10 second-neighbour parameters of the Born-von Karman model. This large number of parameters raises a question about the validity of the postulate on next neighbour interactions in a lattice  let alone the descriptive nature of the lattice dynamics.

In \cite{Gusev-FTSH-RJMP2016} we used these experimental data to derive thermal properties of grey tin. The implied slowest velocity of sound was derived as $v_5 \approx 1.68\cdot 10^3\, m/s$, then, the characteristic parameter $\alpha_0 \approx 1.73$ (denoted $\theta$ in \cite{Gusev-FTSH-RJMP2016}) for silicon and germanium was assumed to be  valid for $\alpha$-tin as well. These numbers together with the lattice constant gave the characteristic temperature $T_0\approx 11$ K.  

These values for $\alpha_0$ and $T_0$ were called predictions, but it turned out they were postdictions, because we overlooked the old work \cite{Hill-PhiMag1952}, where the specific heat of grey tin was studied  at temperature from 7 to 100 K. That paper contains a small set of experimental data, which is reproduced in Table~\ref{greytin1}, where $C$ is in the units $J/(K \cdot mol)$ and temperature is in K.
\begin{table}[!ht]
\caption{Specific heat of grey tin, Ref.~\cite{Hill-PhiMag1952}}
\label{greytin1}
\begin{tabular}{lllllllll}
\hline
$T$ & 7.0 & 8.0 & 9.0 & 10.0 & 12.0 & 15.0 & 20.0 & 25.0 \\
$C$  & 0.180  & 0.301 & 0.464 & 0.674 & 1.18 & 2.08 & 3.74 & 5.31 \\
$T$ & 30.0 & 40.0 & 50.0 & 60.0 & 70.0 & 80.0 & 90.0 & 100.0\\
$C$ &  6.69 & 8.91 & 11.17 & 13.4 & 15.5 & 16.9 & 18.2 & 19.5 \\
\hline
\end{tabular}
\end{table}
Let us now use these data to test our theory. Hill and Parkinson \cite{Hill-PhiMag1952} had two different samples of the 'coarse powder' of grey tin. They measured the specific heat of a higher purity sample from 2 to 20 K. As clear from Table \ref{greytin1} and the corresponding curve, these measurements agree well with the higher temperature measurement, from 12 to 120 K, performed with a poorer quality sample. Our analysis requires the lower temperature set of \cite{Hill-PhiMag1952} which is, unfortunately, rather scarce.

\begin{figure}[ht]
  \caption{QLT regime of grey tin, Ref. \cite{Hill-PhiMag1952}}
  \centering
\includegraphics[scale=0.4]{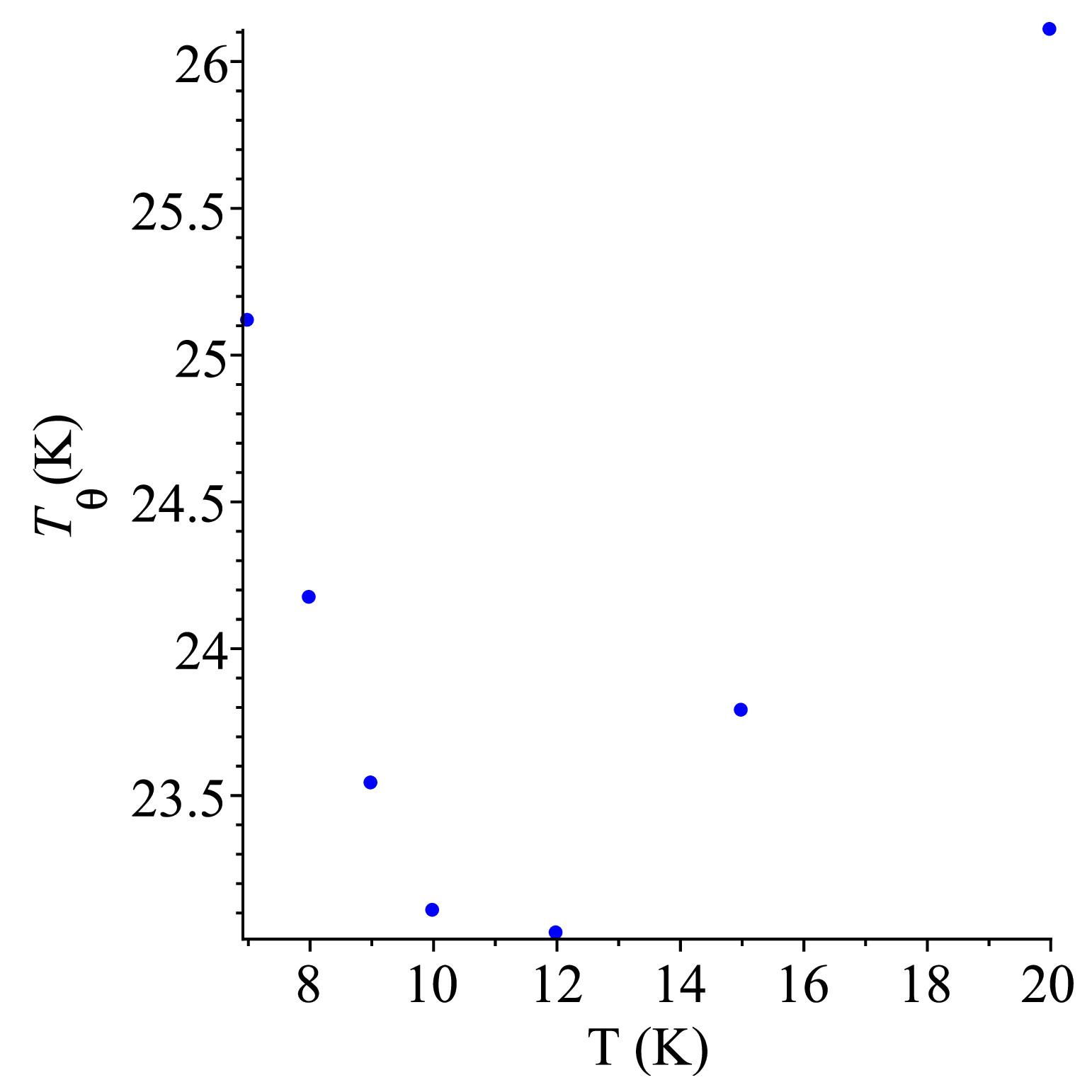}
\label{grey-tin-T3-QLTB}
\end{figure}
We can use either the graph of $C/T^3$ vs $T$ or the plot of the {\em pseudo-Debye temperature} (\ref{thetaT}) to locate the characteristic temperature, $T_0$. If extracted from Fig.\ref{grey-tin-T3-QLTB} of its data, the lowest value of $T_{\theta}$ is 12 K. However, judging from the shapes of similar curves for other materials, true $T_0$ could be between 10 and 12 K, so, it is reasonable take the midpoint of 11 K. This would make a perfect coincidence with the  value predicted in \cite{Gusev-FTSH-RJMP2016}, but it could be, of course, questioned. Nevertheless, 11 or 12 K  is a close match any  way, considering the large experimental uncertainty. The corresponding value of the  dimensionless parameter $\alpha_0=1.73$ automatically agrees with the predicted one and it coincides with the  value for InSb in Table \ref{greytin1}. This  further confirms the similarity of thermal-elastic properties of these two materials. There is no data point for T=11 K in \cite{Hill-PhiMag1952}, the average of two neighbouring points, 10 and 12 K, would give $0.92\ J/(K mol)$, but we prefer to take this value from InSb again, $C_0=1.0$. This is another conjecture about the grey tin's specific heat for future experiments.

Looking at the left, ascending branch of the graph in Fig. \ref{grey-tin-T3-QLTB}, we test the statistical hypothesis (albeit only with 4 data points), whether the corresponding data for $C/T^3$ vs $T$ are governed by the function (\ref{linear}), or equivalently whether $C$ in Table \ref{greytin1} for $T < T_0$ is governed by (\ref{quartic}). Fitting of Eq. (\ref{linear}) to the data in Table \ref{greytin1} gives the values, $d_0 = 1.84(37) \cdot 10^{-4} \ J/(K^4 \cdot mol)$ and $d_1 = 4.96(43) \cdot 10^{-5} \ J/(K^5 \cdot mol)$. Needless to say, that with so few points statistical significance of the fit by a straight line is nearly perfect, with the $\chi^2$-statistic close to zero and the $p$-value almost unity.  This straight line (\ref{linear}) does not cross the axis origin, and it certainly should not, for the absolute zero of temperature is absent from our geometric formalism (\ref{beta}). This means that the quartic term is not the only term in Eq. (\ref{linear}) even the QLT regime, let alone the would be complete expression.  By fitting the function (\ref{quartic}) to the data of Table \ref{greytin1}, we obtained values $d_3 = 2.20(4) \cdot 10^{-4} \ J/(K^4 \cdot mol)$ and $d_4 = 4.56(43) \cdot 10^{-5} \ J/(K^5 \cdot mol)$. They agree with the the coefficients of Eq. (\ref{linear}), up to their statistical uncertainty, as expected.

It is disturbing that the elastic moduli of grey tin, measured in neutron scattering experiments \cite{Price-PRB1971}, as reproduced in Table \ref{group4}, are very different from the ones given in standard chemistry reference book \cite{Landolt-Bornstein-book2001-III}. Apparently, the reference values which were  calculated according to theoretical models should be considered {\em erroneous} in view of the experimental data \cite{Price-PRB1971} and their consistency with the thermal properties discussed above.
 
In Table \ref{group4}, it is striking to see that mechanical (spatial and elastic) characteristics of $\alpha$-tin  crystals almost exactly match the corresponding values for indium antimonide, InSb. This remarkable fact was first noted  in \cite{Price-PRB1971}. This is an instance of the phenomena of scaling explored in Sect. \ref{scaling}. However, this match offers a unique opportunity to verify the most special feature of the field theory of specific heat: the would be measured specific heat of $\alpha$-Sn should coincide with the corresponding function for InSb \cite{Price-PRB1971}. This is our ultimate goal -  to create a theory that can describe thermal properties of many materials at once, and by doing so to predict the ones not measured yet. In the literature, the transition from $\beta$-Sn to $\alpha$-Sn is described as a way to produce grey tin. However, all references which we found, e.g. cited in \cite{Houben-PRB2019}, describe the opposite transformation, from the diamond lattice to the metallic tin, only in thin films of $\alpha$-Sn deposited on zincblend substrates.

This raises the question, if melting of grey tin can occur in principle? 
The work \cite{Abrikosov-Calphad2019} reported the specific heat of $\beta$-tin collected from different sources from 20.54 up to 300 K.  This raises a question, if the metallic form of tin could be studied below the {$\alpha$-$\beta$-transition}, why the diamond lattice tin would not be studied {\em above} this transition's temperature?  Indeed, the only data point determined at high temperature of 283.6 K,  $C= 25.44\ J/(K\cdot mol)$, was measured nearly a century ago \cite{Lange-ZPCL1924}, as reproduced in Table V of Ref. \cite{Dayal-PIASA1941}, p. 481. It is  known that crystalline seeds of $\alpha$-tin significantly change this transition \cite{Zeng-CGD2015}. This lattice transformation also occurs at different temperatures for  thin films. By employing the similarity of grey tin and indium antimonide we conjecture, that the proposed Dulong-Petit value in Table \ref{group4} really corresponds to the {\em melting} temperature of grey tin, which could be achieved for very pure single crystals isolated from other lattice phase. If this phenomenon could occur, it would likely occur at temperature close to the melting temperature of InSb,  $T_m =800\ K$. These are different chemical substances, but if our hypothesis that thermal phenomena, including phase transitions, are governed by elastic properties of matter, this conjecture is viable and worth exploring experimentally.

New precision measurements of the specific heat of grey tin could deliver more numerical values to check the proposed functional form of $C$.  The method of growing single crystals of grey tin used to be difficult \cite{Ewald-JAP1958}, but a simpler method is known now \cite{Styrkas-IM2005} that can  produce even shaped single crystals. We hope that this situation with the scarce  data \cite{Hill-PhiMag1952} would encourage experimentalists to measure the thermal properties $\alpha$-tin. Such experiments are urgently  needed for the reference texts used in experimental physics and technology.

\subsection{Revisiting the natural diamond data}  \label{naturaldiamond}

It was a sheer coincidence that the first set of high precision data of specific heat we found, were those of  the group IV elements, \cite{Desnoyers-PhilMag1958,Flubacher-PM1959}. These data gave an opportunity to test the new theoretical  proposal. In Ref. \cite{Gusev-FTSH-RJMP2016} preliminary statistical estimates of the characteristics of specific heat of diamond, silicon and germanium were made. There, we concluded that the data for natural diamond of Ref. \cite{Desnoyers-PhilMag1958} (the measurements of were done with 160 g of industrial quality diamonds with the average dimension of 3 mm) did not allow us to make a statistically significant selection between the two powers, $T^3$ vs. $T^4$. However, we have re-analyzed these data with the extended statistical fit (\ref{quartic}) proposed in Ref. \cite{Gusev-QLTBSH-RSOS2019}, and our conclusion is  different now: the quartic power law clearly gives a contribution, in the quasi-low temperature range, from 80 to 130 K. A combination of two factors led to the earlier wrong statement: 1) the empirical choice of the temperature range which was too long (from 27 to 174 K); 2) a single quartic term  in the fitting equation was used instead of Eq. (\ref{quartic}). This once again shows that the surface specific heat must be always taken in account when analysing data in the QLT regime.

\begin{figure}[ht]
  \caption{QLT regime of natural diamonds, Ref. \cite{Desnoyers-PhilMag1958}}
  \centering
\includegraphics[scale=0.4]{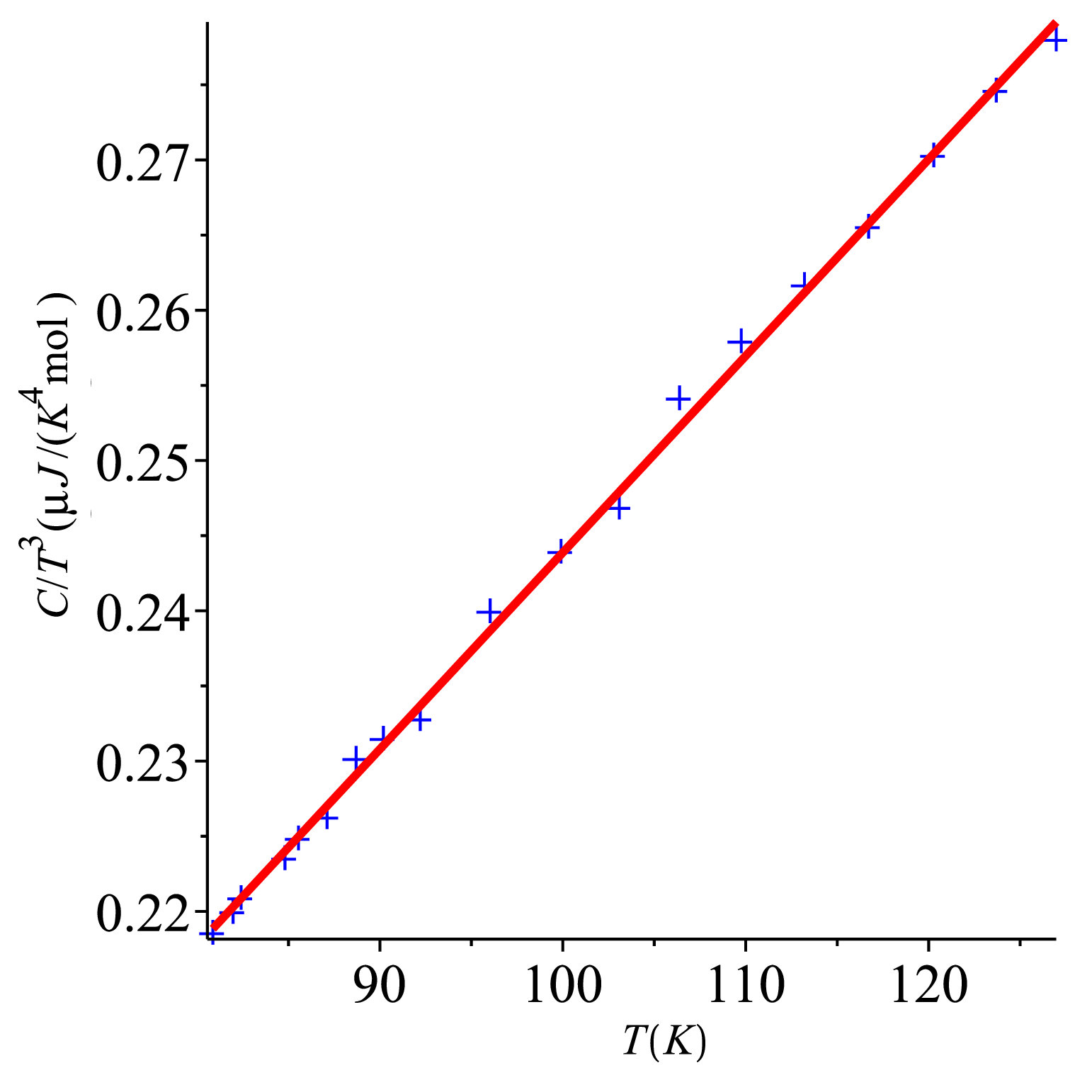}
\label{diamond-Morrison-T}
\end{figure}

As one can see, the graph of $C/T^3$ vs $T$ for natural diamond in Fig. \ref{diamond-Morrison-T} is as good as other similar plots. The ansatz (\ref{quartic}) is fitted to the raw data of Ref. \cite{Desnoyers-PhilMag1958} with the coefficients $d_3=1.155(16) \cdot 10^{-7} \ J/(K^4 \cdot mol)$ and $d_4=1.286(15) \cdot 10^{-9} \ J/(K^5 \cdot mol)$. The linear fit (\ref{linear}) of $C/T^3$ gives similar values, $d_0=1.130(13) \cdot 10^{-7} \ J/(K^4 \cdot mol)$ and $d_1=1.308(13) \cdot 10^{-9} \ J/(K^5 \cdot mol)$. These sets of values agree with each other, within standard errors. This shows good statistical significance of the hypothesis (\ref{quartic}). It is obvious that the graph in Fig. \ref{diamond-Morrison-T} is nearly a straight line.
 These coefficients also show that at about 100 K both terms, cubic and quartic, give contributions to the total specific heat, with comparable absolute values. It means that the surface specific heat should always be taken into account (see also Sect. \ref{diamondsize}). The complete field theory of specific heat should be calibrated, which is a task for future.

We confirm the above conclusion with the higher precision data, reported (but not published) in Ref. \cite{Cardona-SSC2005}. The work  \cite{Cardona-SSC2005} produced an excellent data set, which has  368 data points for $C$ between 28.26 and 280.30 K. This study explored the dependence of diamond's specific heat on isotope content for three different isotope combinations. The technology of producing artificial diamonds introduces iron atoms up 0.15 mol\% of diamond, and this contamination can  change diamond's thermal properties \cite{Cardona-SSC2005}. Therefore, we refrain here from using artificial diamond data and leave the isotope dependence for a later study.  

The diamond's thermal characteristics deviated from those for other two elements, silicon and germanium\cite{Gusev-FTSH-RJMP2016}. Specifically, the characteristic temperature was determined as 173.3 K, which gave the value $\alpha_0=1.44$, different from 1.73 for the other elements. In Discussion section of \cite{Gusev-FTSH-RJMP2016}, it was conjectured that the low temperature thermal properties of diamond should really be similar to those of other elements with the diamond type lattice.In fact, diamond has the widest known range of the quasi-low temperature regime among all chemical elements (this range is probably comparable to zinc-blend compounds with similar mechanical properties, e.g. cubic boron arsenide (BAs) \cite{Tian-APL2019}). This wide  temperature range makes it difficult to assign a specific value to $T_0$ (as seen in Fig. \ref{diamond-T3-size} of Sect. \ref{diamondsize}). Instead of leaving the old value of  $T_0$, we decided to trust the scaling emerging in the data and conjectured that it could really be determined via the specific heat $C_0$. Looking at its values for the carbon group elements in Table \ref{group4} we extrapolated this sequence for diamond. Then, the characteristic temperature could be found from the data of specific heat in \cite{Desnoyers-PhilMag1958}, it came out $T_0 = 161\ K$. The new $T_0$ produces the new dimensionless parameter $\alpha_0= 1.55$, which seems to continue the decreasing trend along the element's column.

\begin{figure}[ht]
  \caption{Specific heat of a 48.1 mg natural diamond, unpublished data for Ref. \cite{Cardona-SSC2005}}
  \centering
\includegraphics[scale=0.5]{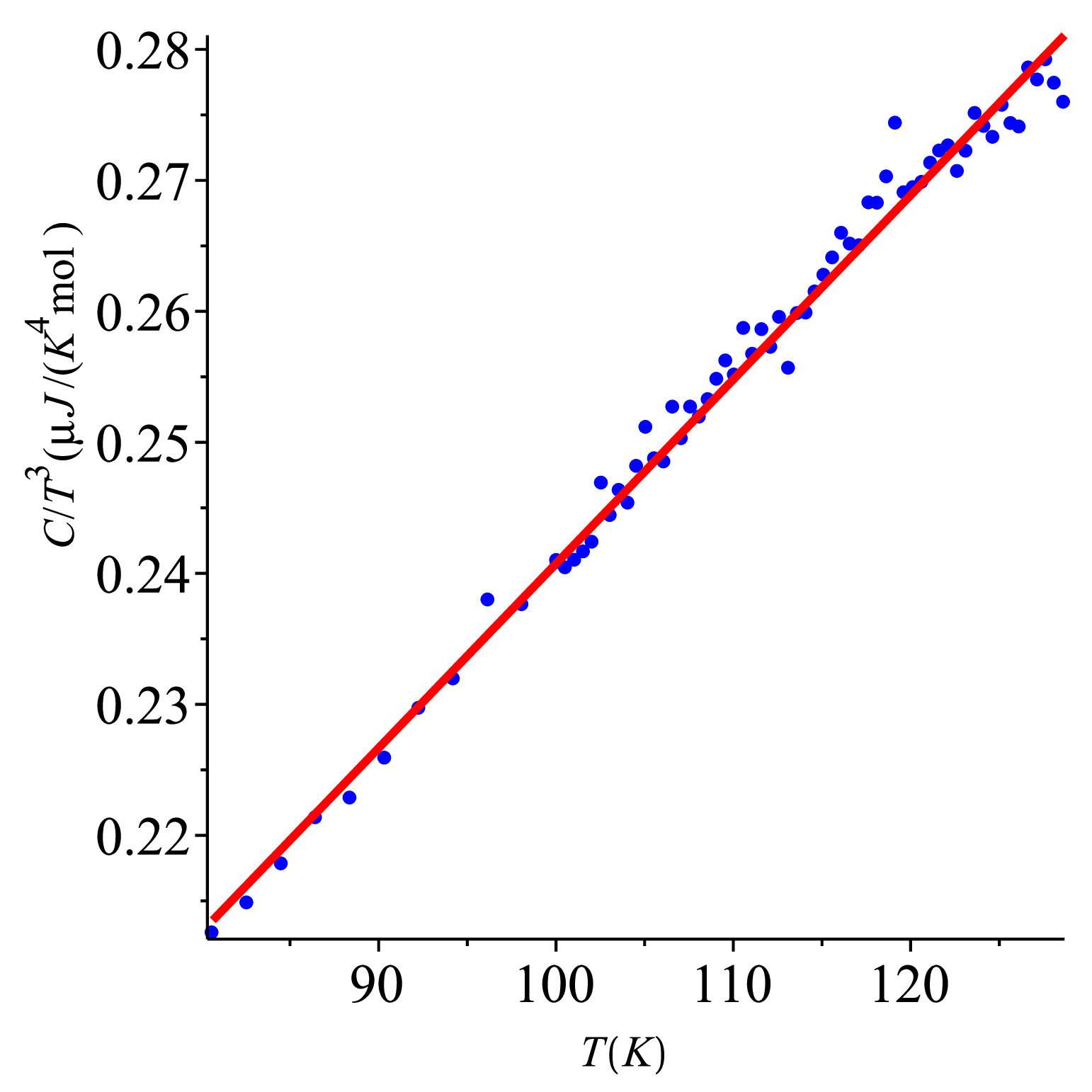}
\label{diamond-Cardona-T4}
\end{figure}

The ansatz (\ref{linear}) is fitted to the data of \cite{Cardona-SSC2005}, which are presented in Fig. \ref{diamond-Cardona-T4},
 with the coefficients $d_0=1.001(21) \cdot 10^{-7} \ J/(K^4 \cdot mol)$ and $d_1=1.407(19) \cdot 10^{-9} \ J/(K^5 \cdot mol)$, while fitting (\ref{quartic}) to the specific heat $C$, recovered from $C/T^3$, gives  $d_3=1.070(30) \cdot 10^{-7} \ J/(K^4 \cdot mol)$ and $d_4=1.346(26) \cdot 10^{-9} \ J/(K^5 \cdot mol)$, for a gem quality natural diamond of mass 48.1 mg (0.24 carats). Using this mass and the diamond density, it is easy to estimate the average dimension of a stone. For an uncut stone it is fair to approximate its shape by a sphere, which gives a value of about 3 mm, same as for the diamonds used in Morrison's study \cite{Desnoyers-PhilMag1958}. This means that their effective sizes (specific surfaces) were comparable, and this fact is reflected in the values $d_0$ ($d_3$) which are close. Without more detailed information about the samples of each work \cite{Desnoyers-PhilMag1958,Cardona-SSC2005}, we cannot investigate whether an apparent difference in these coefficients is caused by the difference of effective sizes or experimental uncertainties.

\begin{figure}[ht]
  \caption{QLT behaviour of the specific heat of natural diamond, unpublished data for Ref. \cite{Cardona-SSC2005}}
  \centering
\includegraphics[scale=0.5]{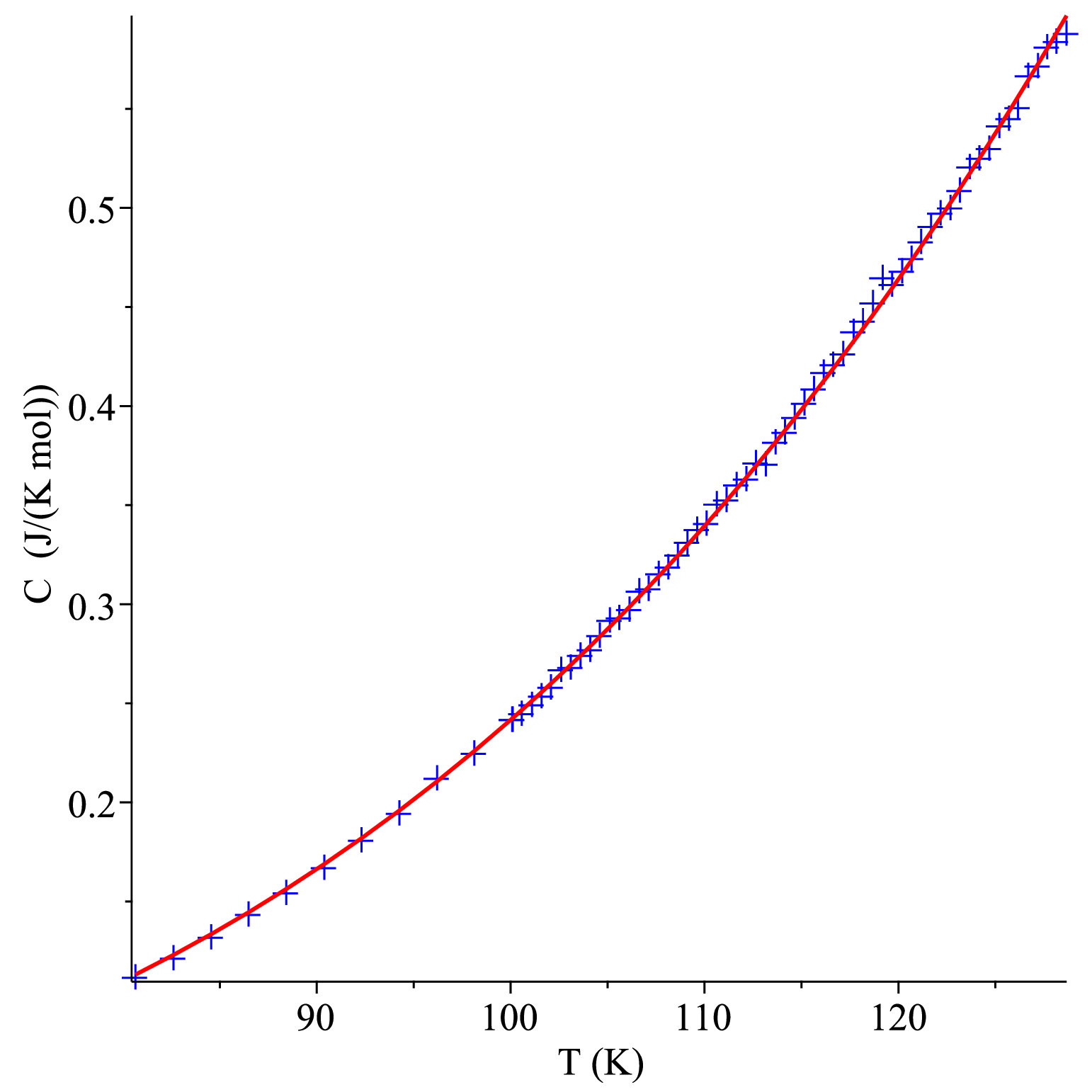}
\label{diamond-Cardona-QLT}
\end{figure}

Summarizing, two different data sets, obtained at two very different labs, with different samples, produced very close values that characterize the specific heat of natural diamond, including the surface specific heat. The result of fitting the data shown in Fig. \ref{diamond-Cardona-T4} with (\ref{quartic}) is displayed in Fig. \ref{diamond-Cardona-QLT}. 

Natural diamond was the first material whose specific heat was used to test the Einstein's model based on the Planck's quantum theory \cite{Einstein-AdP1907}. It was the worst possible choice though, because of the diamond's high temperature of the QLT threshold. The specific heat of diamond was not measured with needed precision for another forty years, and even after that it was confusing to determine a right power combination, as the present paper and \cite{Gusev-FTSH-RJMP2016} show. 

The problem of the $\alpha_0$-parameter of natural diamond remains, one contributing factor is the large surface heat capacity of the diamond samples, as discussed in Set. \ref{diamondsize}, which is less significant for the silicon and germanium measurements considered in \cite{Gusev-FTSH-RJMP2016}. Besides, the elastic properties of diamond seem to be  different from other elements of group IV. Namely, silicon \cite{Middelmann-PRB2015} and germanium \cite{Barron-AiP1980} have negative thermal expansion coefficients (contraction of a crystal) at some low temperatures. However, no negative thermal expansion, i.e. thermal contraction, was detected in experiments with diamond \cite{Stoupin-PRL2010}. In our model, thermal expansion is important to take into account \cite{Gusev-QLTBSH-RSOS2019} because it changes the volume and the lattice constant, which enters the dimensionless variable $\alpha$.

\subsection{Scaling in the specific heat data of zincblend compounds} 
\label{calibration}

The phenomenon of scaling has been studied empirically for a long time in the physics of glasses (see some refs. in \cite{Gusev-QLTBSH-RSOS2019}). The same phenomenon occurs in crystalline matter, but it is less spectacular because of anisotropy and larger surface heat due to samples sizes. The phenomenon of scaling can be exposed by the parameter $\alpha_0$, which should be (almost) the same for all materials within the lattice class. This conjecture is not proven for the elements, but it is better supported by the zincblend compounds as seen in Table \ref{group4}. 

The calibration of the geometrical theory of specific heat \cite{Gusev-FTSH-RJMP2016} requires two experimental parameters, $A$ and $B$. Parameter $A$ determines the scaling in a vertical direction (the 'height') of the specific heat graphs for a given group of materials. In general, it could be fixed by the Dulong-Petit value, $C_{DP}$, at the phase transition temperature, but $C_{DP}$ is conjectured to be the same for all materials in the same lattice class. Another special value of specific heat is $C_0$ at the characteristic temperature $T_0$, which indeed is the same for the zincblend lattices and nearly the same for the diamond lattices in Table \ref{group4}. Parameter $B$ , which governs the horizontal scaling (the 'width'), can be determined with the characteristic temperature, $T_0$. Therefore, according the revised Debye scaling hypothesis which we adopt here, specific heat functions in the same lattice group are scaled by $T_0$, i.e.  if  temperature is made dimensionless as $\tau =T/T_0$, the functions should coincide, i.e. form the universal curve ('master curve' as called in the physics of glasses). 

\begin{figure}[ht]
  \caption{QLT regime of specific heat of diamond and zincblend lattice materials}
  \centering
\includegraphics[scale=0.5]{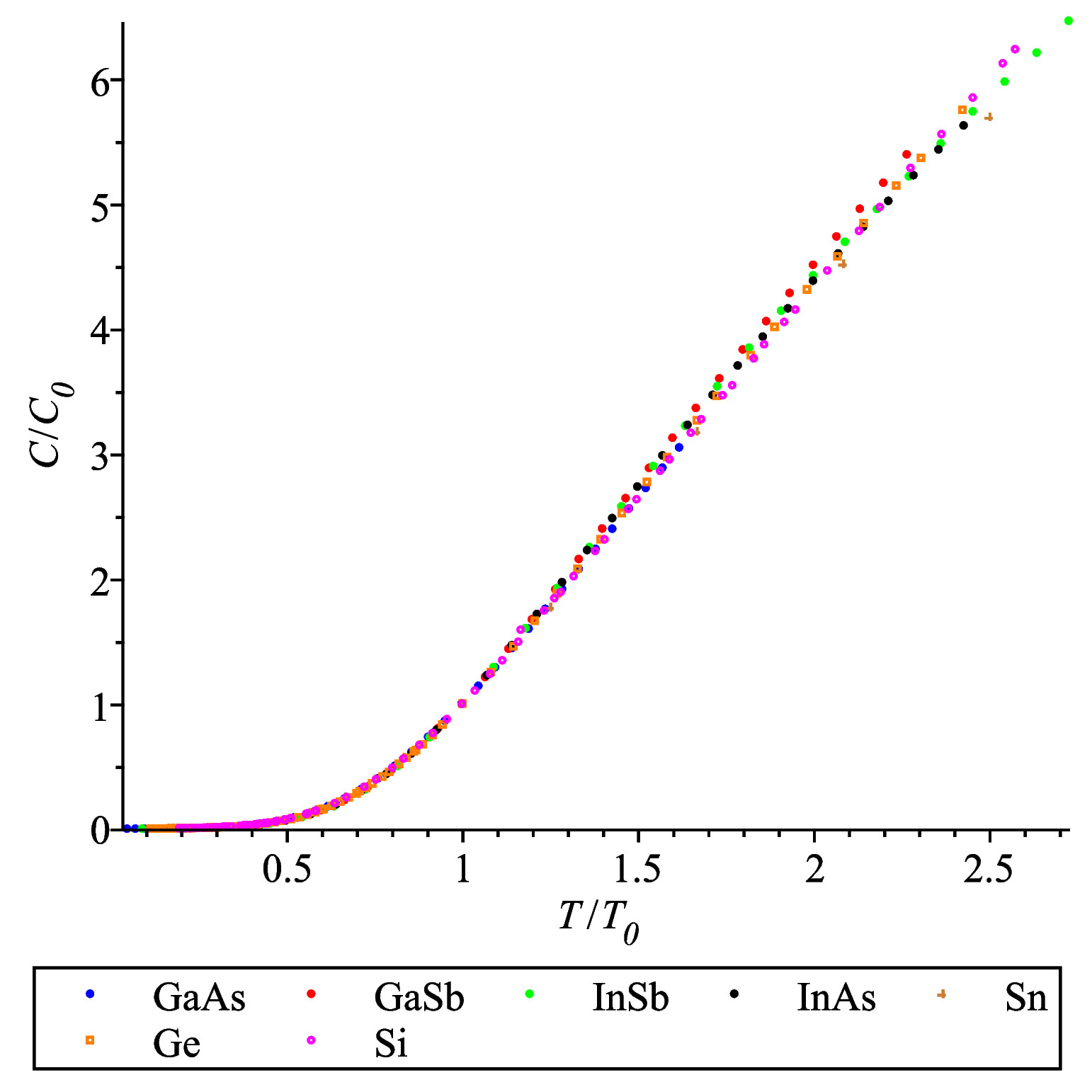}
\label{scaling-zincblend}
\end{figure}

\begin{figure}[ht]
  \caption{QLT regime of specific heat of diamond and zincblend lattice materials}
  \centering
\includegraphics[scale=0.5]{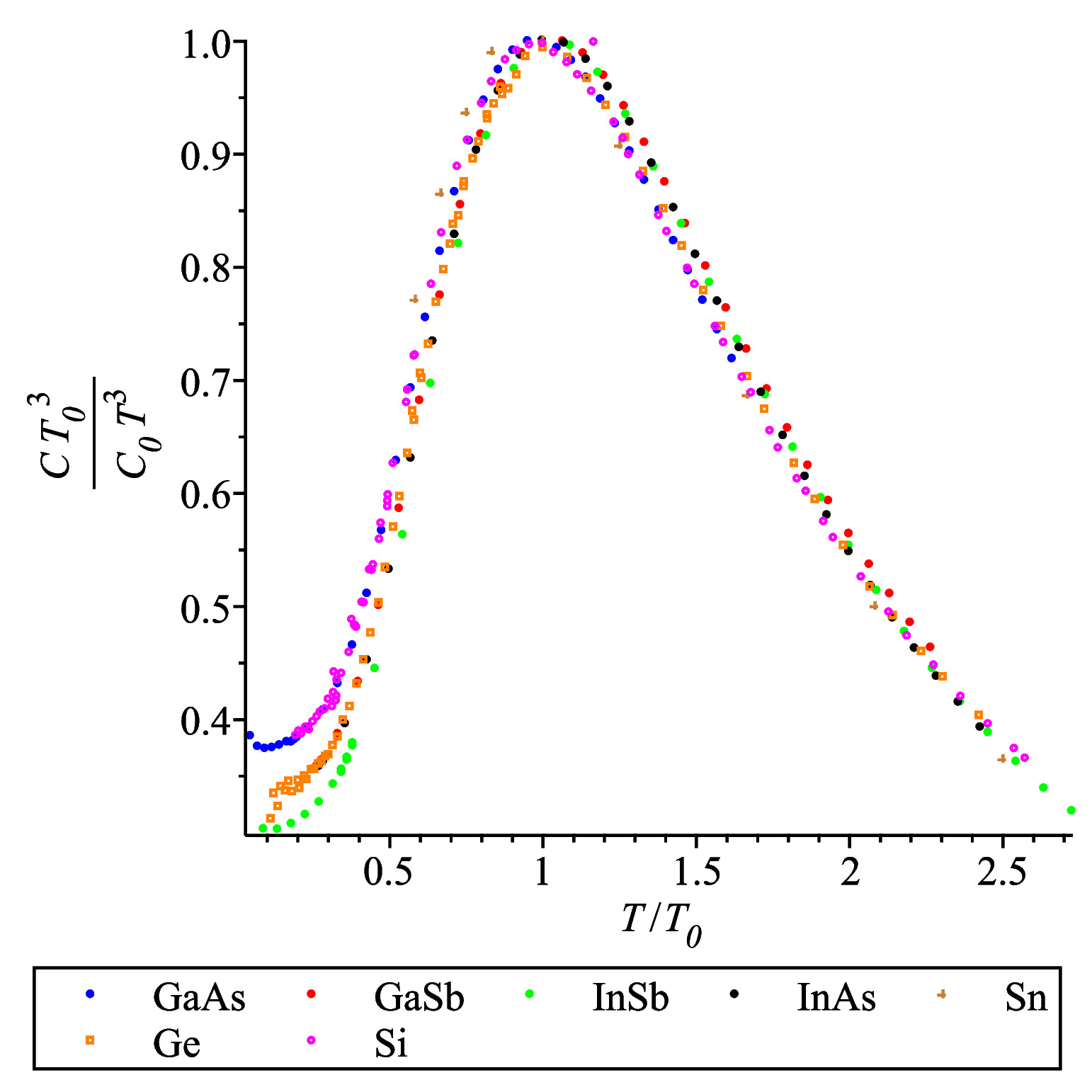}
\label{scaling-T3-zincblend}
\end{figure}

We introduce the scaled parameter,
\begin{equation}
\tilde{d}_1= d_1 \cdot {T_0}^4,
\end{equation}
(the dimensionality of $d_1$ is $J/(K^5 \cdot mol)$). As seen from Table \ref{group4}, the scaled coefficients $\tilde{d}_1$ for all diamond lattice materials nearly coincide. This fact is certainly not accidental, it is a feature of the scaling phenomena: specific heat functions of all materials in the same crystal class should be the same function, and the material specific characteristic, in this case $T_0$, allows to make up a material specific function. Because the range of temperatures, at which the specific heats of zincblend compounds are measured, vary we cannot produce the master curve at high temperature. It looks convincing on the graphs of specific heats, scaled with the characteristic values, like in Figs. \ref{scaling-zincblend} and \ref{scaling-T3-zincblend}.
We should note that the scaling here cannot be perfect in principle because of the surface specific heat which depend of the sizes and shapes of samples. In Figs. \ref{scaling-zincblend} and \ref{scaling-T3-zincblend}  the surface heat capacity which behaves as $T^3$ at low temperatures is visible as the left 'tails' of different magnitudes. The best coincidence is observed for a set of zincblend compounds of Ref. \cite{Cetas-PR1968} because that work used large crystals and therefore its measurements are closer to the so-called 'bulk' specific heat. 

Even though we expect the same $C_0$ for all materials of the same lattice class, the specific heats in these figures are divided by their $C_0$s,  because experimental and statistical uncertainties introduce some dispersion, which is really not intrinsic. In any case, the visualization is only indicative, and it should not be entirely relied upon in physics development.

\section{Experimental evidence of the surface heat capacity}  \label{surface}

The problem of  the surface heat capacity of solid matter was attacked in 1950-60s, but experimental results did not  match theoretical predictions \cite{Morrison-JCE1957}. Amidst confusion and contradiction, no further dedicated experiments have been performed, to our knowledge. Below we re-analyze some relevant old works with published experimental data and demonstrate that the surface heat capacity is much easier to observe than it was previously believed. 

We found that a surface contribution to the total specific heat is proportional to the effective size of a condensed matter body, in agreement with our theoretical proposal and in contradiction to the old theories. This contribution is present at any temperature, and at some low temperatures, it becomes dominating as the cubic in temperature term, which depends on the body's size and shape, i.e. not universal (not 'bulk').  Thus, any sample of condensed matter possess, at any temperature, the surface heat capacity, which is {\em relatively} higher at some sufficiently low temperature. The so-called 'bulk' specific heat can only extracted from measurements if samples of different sizes were studied.

\subsection{Surface specific heat of the sodium chloride  powder}  \label{NaClpowder}

Among a few dedicated measurements is Ref. \cite{Barkman-JCP1965}, which contains a table of the specific heat of rock salt (sodium chloride), from 4 to 20 K. This work used fine grains (powder) of $NaCl$, with the {\em specific surfaces} 41.8 and 78.3 $m^2/g$. A specific surface is equal to the total surface area of all particles divided by their total mass. It is experimentally determined by the absorption of gas on surfaces. Therefore, it gives no information about the  distribution of particles' sizes, which can only be found directly. Yet, the specific surface can serve as a proxy for the inverse of the effective size of a sample, $r$,  (\ref{effsize}). Since the density $\rho$ is a constant, the specific surface is a quantity that describe the linear average size of particles,
\begin{equation}
s = S/m=S/(V \rho) = 1/(r \rho).
\label{ssurface}
\end{equation}
thus, it is suitable for testing the surface hypothesis of our theory (\ref{2D}).

\begin{figure}[ht]
  \caption{Specific heat of sodium chloride powders, Refs. \cite{Barkman-JCP1965,Barron-PRSA1964}}
  \centering
\includegraphics[scale=0.5]{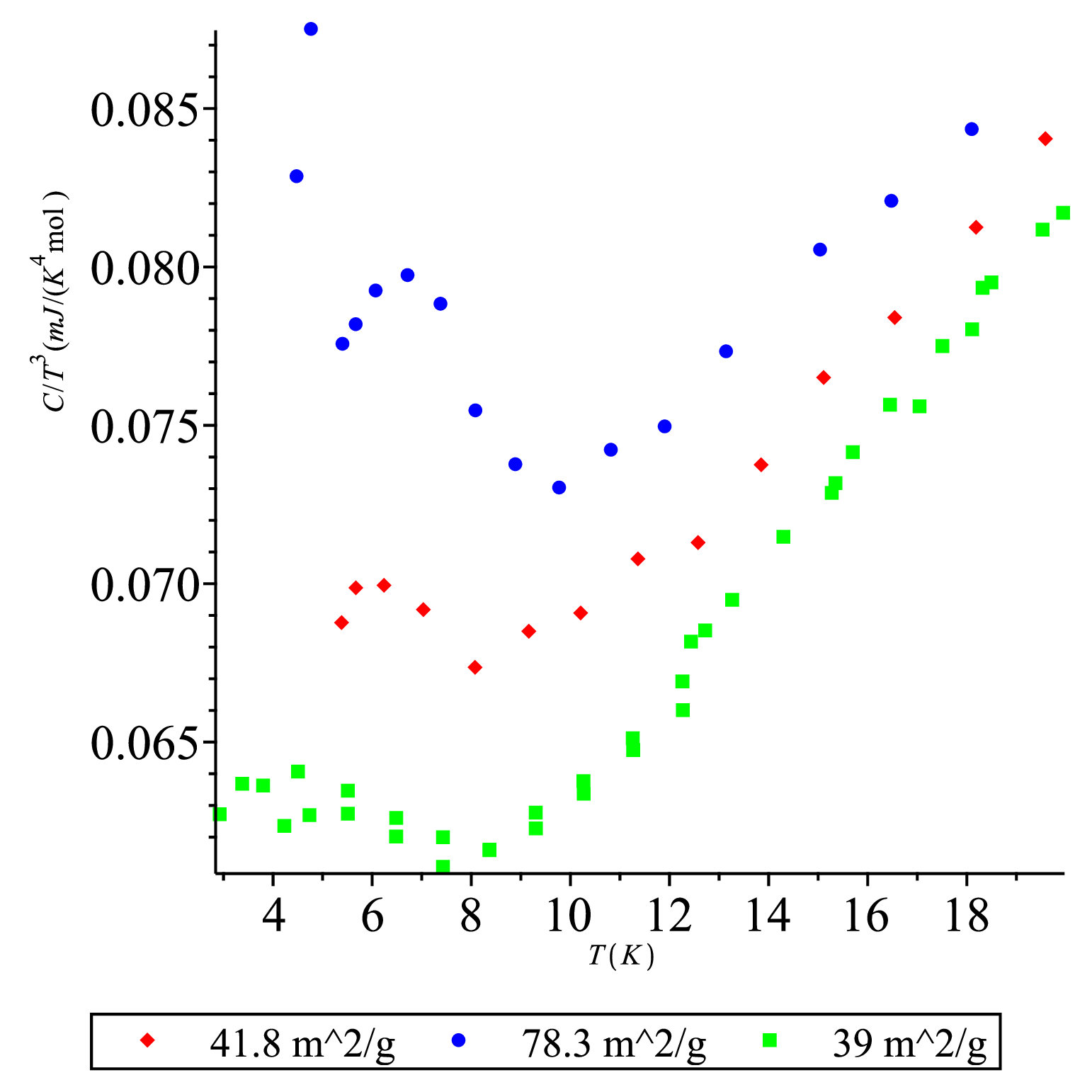}
\label{NaCl-T3-size}
\end{figure}

Let us look at the data of \cite{Barkman-JCP1965} in Fig. \ref{NaCl-T3-size}. 
The sample with the specific surface $41.8 \ m^2/g$, in the range of temperature from 12.6 to 19.6 K, can be fit by Eq. (\ref{quartic}) with the coefficients $d_3= 4.91(20) \cdot 10^{-5} \ J/(K^4 \cdot mol) $ and  $d_4= 1.78(4) \cdot 10^{-6}  \ J/(K^4 \cdot mol)$. The sample with the specific surface $78.3 \ m^2/g$, in the range of temperature from 11.9 to 18.1 K, can be fit by Eq. (\ref{quartic}) with the coefficients $d_3= 5.75(39) \cdot 10^{-5} \ J/(K^4 \cdot mol) $ and  $d_4= 1.49(22) \cdot 10^{-6}  \ J/(K^4 \cdot mol)$. Therefore, the $d_4$ coefficients overlaps, within their standard errors, while the coefficients $d_3$ do not, i.e. the slopes of the graphs in Fig. \ref{NaCl-T3-size} can be statistically considered as equal, while their intersections with the vertical axis are different, as evident from the vertical displacement of the graphs. The ratio of the effective sizes for two powders is $78.3/41.8 \approx 1.87$, while the indirect estimate of the surface specific heat by the QLT data via $d_3$ values is $5.75/4.91 \approx 1.17$, which is at least of the same order of magnitude; further analysis is limited by scarce data and high experimental uncertainties.

We can add the data for NaCl powder with the specific surface $39\ m^2/g$ (average particle size 70 nm) from Ref. \cite{Barron-PRSA1964}, which presented the results of \cite{Morrison-PFS1956}. Its graph  in Fig. \ref{NaCl-T3-size} has a  slope comparable to Barkman's samples and located below both graphs of data from \cite{Barkman-JCP1965} as expected. Below 8 K this data set clearly displays the $T^3$ behaviour, which is quantitatively quite different from two other sets. The analysis of the of the data from \cite{Barron-PRSA1964} in the quasi-low temperature region, below the characteristic temperature $T_0 = 26.8\ K$, can describe the specific heat by the quartic function (\ref{quartic}), the (universal) coefficient $d_4 = 1.85(8)\cdot 10^{-6}\ J/(K^4 \cdot mol)$ and the (size-dependent) coefficient $d_3 = 4.49(14) \cdot 10^{-5}\ J/(K^4 \cdot mol) $. These values show that the quartic contribution can be viewed as almost the same (universal), while the cubic one is different for different specific surfaces, as clear from Fig. \ref{NaCl-T3-size}. Specific heat  of the true 'bulk' sample was also measured by Morrison's group, but the data were never published. Let us note that above date sets were also analysed from the point of view of traditional Debye theory, which predicts the $T^2$ contribution from the surface heat capacity \cite{Onsager-JCP1960}. Because this theoretical study is erroneous, its phenomenological analysis could not find an evidence of the sought functional behaviour in the data \cite{Onsager-JCP1960}, p. 1461.

Another study of the sodium chloride powders was done  in the Morrison's lab \cite{Morrison-CJC1955}. This work measured specific heat of the Na Cl powder produced by vaporization, with the specific surfaces, $s$, 38 and 59 $m^2/g$, from 9 to 21 K. The experimental data were not published, and the results were presented in Fig. 1 on p. 242, \cite{Morrison-CJC1955}, as graphs of the Debye temperature. Unfortunately, due to specific technical problems, authors did not measure the specific heat below 9 K, which happened to coincide with the onset of the dominating behaviour of the surface heat capacity. 

\subsection{Surface specific heat of {$\alpha$-Sn}}  \label{tinpowder}

It was quite intriguing to find another  set for the specific heat data of the powder of grey tin at even lower temperature that was obtained at a different lab \cite{Webb-PRSA1955}, shortly after the work \cite{Hill-PhiMag1952}. It is a very small number of data points, which are reproduced in Table \ref{greytin2} (units $mJ/(K \cdot mol)$).
\begin{table}[!ht]
\caption{Specific heat of grey tin, Ref. \cite{Webb-PRSA1955}}
\label{greytin2}
\begin{tabular}{lllllll}
\hline
$T (K)$ & 1.5 & 2.0 & 2.5 & 3.0 & 3.5 & 4.0\\
$C $  & 0.69  & 1.63 & 3.21 & 5.55 & 9.06 & 14.9 \\
$C/T^3$  & 0.204  & 0.204 & 0.205 & 0.206 & 0.211  & 0.233 \\
\hline
\end{tabular}
\end{table}

This data set is remarkable because it displays nearly a cubic law below 3 K, as obvious from the third row of Table \ref{greytin2}, where $C/T^3$ is presented. From the statistical analysis of the first data set, Table \ref{greytin1}, Ref. \cite{Hill-PhiMag1952}, we found that the cubic contribution in (\ref{quartic}) is characterized by the parameter $d_3= 2.20(4) \cdot 10^{-4} \ J/(K^4 \cdot mol) $ or  $d_0= 1.84(37) \cdot 10^{-4}  \ J/(K^4 \cdot mol)$  (for the raw data or $C/T^3$ correspondingly).  These parameters agree, up to experimental and statistical uncertainties, with the average $2.11 \cdot 10^{-4} J/(K^4 \cdot mol)$ taken for the third row of Table \ref{greytin2}. The direct  fit of  Eq. (\ref{quartic}) to these data gives values $d_3= 2.37(58) \cdot 10^{-4} \ J/(K^4 \cdot mol) $ and  $d_4= 1.36(22) \cdot 10^{-3}  \ J/(K^4 \cdot mol)$. Clearly, the $T^4$ cannot properly determined with two data points, and the cubic power coefficient is almost the same as the one obtained by averaging.
The consistency of these coefficients with with the higher temperature data set is encouraging, provided the effective sizes of the powders' particles in both works were comparable.

\subsection{Surface specific heat of natural diamonds}  \label{diamondsize}

Long time ago, the Morrison's lab discovered a peculiar property of the specific heat when exploring the dependence on the particles sizes.  The work \cite{Dugdale-PRSA1954} studied powders of rutile, titanium dioxide ($TiO_2$) with two specific surfaces. The authors found that the excess of the heat capacities of samples with small effective sizes (or specific surface for powders) over the 'bulk' value is observed at higher temperatures, up to 270 K. Unfortunately, authors did not publish these experimental data. 

The same phenomenon was observed in their work on sodium chloride powders \cite{Morrison-PFS1956}. They reported the maximum of the excess of the specific heat of small particles over the bulk value at around 40 K. This conclusion was meant to match the theory of Montroll \cite{Montroll-JCP1950,Morrison-PFS1956}. For this section, it is important that the size effect was confirmed by experiment to temperature up to 260 K. The authors did not publish the specific heat data for {\em both} samples of NaCl powders, $39$ and $59\ m^2/g$, of the work \cite{Morrison-PFS1956}.

Thus, the observation of surface heat capacity in all temperature range went  unnoticed. Besides, lattice dynamics theories dictated the surface effect could only be relevant for 'low' temperatures and 'small' particles \cite{Patterson-CJC1955}.  We show here that the size effect is  prominent even for 'macroscopic' bodies at 'higher' temperatures. This means that the assumed 'bulk' specific heat does not really exist, as the surface heat capacity is always present and measured.

\begin{figure}[ht]
  \caption{Specific heat of two natural diamonds, unpublished data for Ref. \cite{Cardona-SSC2005}}
  \centering
\includegraphics[scale=0.5]{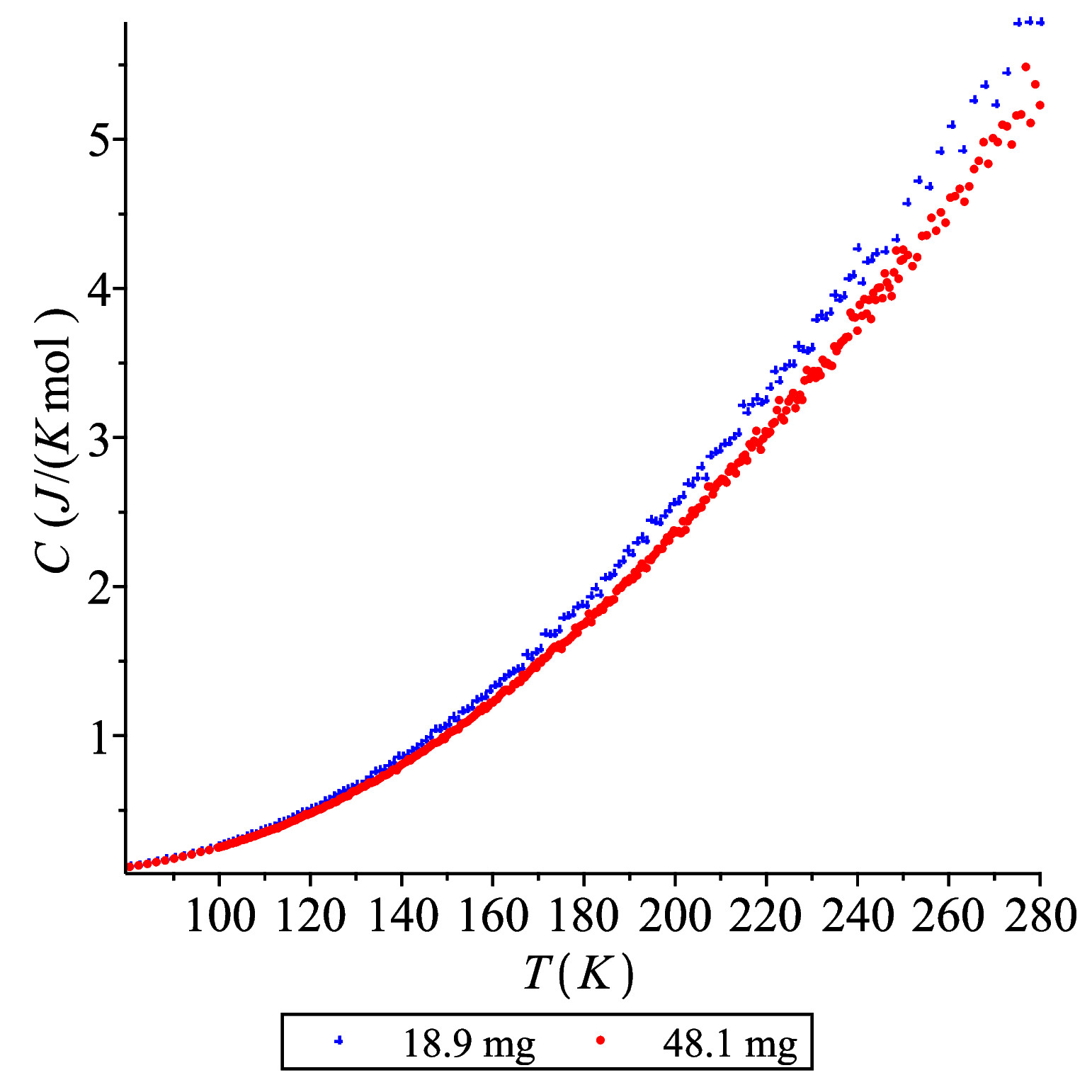}
\label{diamond-T-size}
\end{figure}

\begin{figure}[ht]
  \caption{$C/T^3$ of two natural diamonds, unpublished data for Ref. \cite{Cardona-SSC2005}}
  \centering
\includegraphics[scale=0.5]{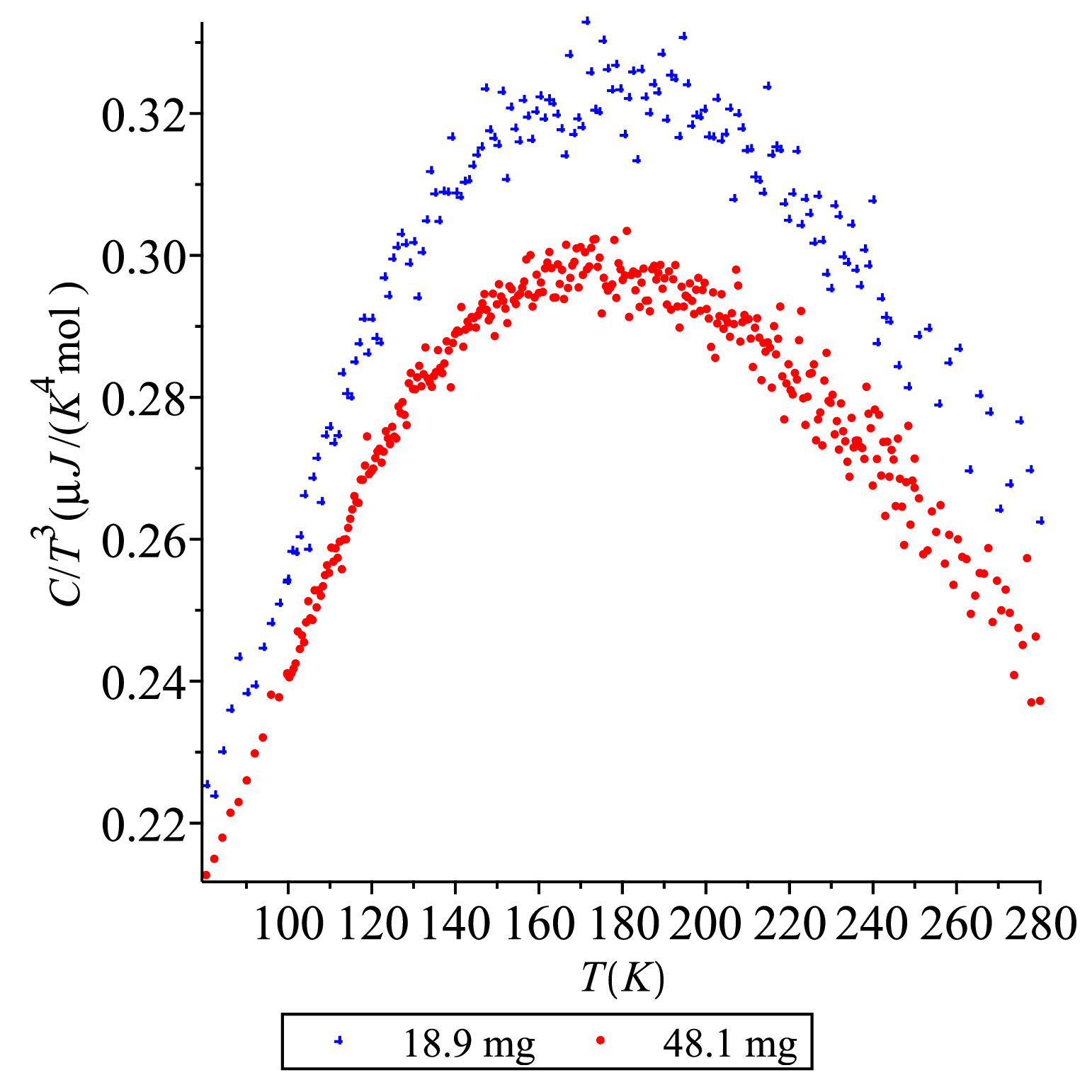}
\label{diamond-T3-size}
\end{figure}

Taking again unpublished data from Ref. \cite{Cardona-SSC2005}, where two natural diamonds of different masses, 48.1  mg and 18.9 mg,  were studied, one can clearly see the specific heat difference due to different specific surfaces in Fig. \ref{diamond-T-size}.
To emphasize this effect in the lower temperature region, we can plot the usual $C/T^3$ vs $T$, Fig. \ref{diamond-T3-size}.

To quantify this surface effect, we take the $d_0$ values for both samples to obtain their ratio as $d_0^{(2)}/d_0^{(1)} = 0.117/0.0910 \approx  1.27$. The ratio of their effective sizes can be estimated from the corresponding volumes, which are proportional to stones' masses, $(m^{(1)}/m^{(2)})^{1/3} = (48.1 / 18.9)^{1/3} \approx 1.37$. This time we found not only the qualitative agreement, but also rather close numerical values, which show that the cubic in temperature contributions are indeed inversely proportional to the effective (linear) sizes. This difference is measured at higher temperatures as well, but our theory is still not complete, so quantitative analysis cannot be done yet.

\section*{Summary}

\begin{itemize}
\item
The previously calculated thermal characteristics of  grey tin are confirmed by available  experimental data, up to experimental and statistical uncertainties.
\item
The specific heat function of $\alpha$-Sn is expected to match the specific heat of InSb, including the melting temperature, when direct measurements would be made with single crystals of pure grey tin.
\item
The specific heat of natural diamond behaves similarly to other diamond lattice materials. Its QLT asymptotic contains the fourth power of temperature, as confirmed by two independent data sets. 
\item
The field theory of specific heat predicts that the non-universal surface specific heat is proportional to the cubic power of temperature and inversely proportional to the sample's effective size. This finding is supported by the data sets of natural diamonds, and of the powders of grey tin and sodium chloride.
\item
The quasi-low temperature behaviour of the specific heat of two-dimensional condensed matter system, e.g. suspended graphene, is predicted to be cubic in temperature, with the quadratic, non-universal contribution from the edges.
\item
The scaling in the specific heat data of zincblend compounds is exhibited by the universality of these functions, scaled by their characteristic temperatures.
\end{itemize}

\section{Discussion}  \label{discussion}

\subsection{Confrontation of  the Debye theory with experiment}

The critical analysis of the Debye theory \cite{Gusev-QLTBSH-RSOS2019}, made {\em after} the results of the field theory of specific heat were axiomatically derived  \cite{Gusev-FTSH-RJMP2016},  showed that this textbook model is theoretically inconsistent and experimentally erroneous. The similar critique of the Debye model was also done by other authors, e.g. \cite{Raman-PIASA1941,Passler-AIP2013}. Its key predictions, the cubic temperature law of the specific heat at 'low' thermodynamic temperatures and the universal limiting value of specific heats at melting temperatures are incorrect.  Experimental data exhibit the linear in temperature growth of specific heat beyond the traditional Dulong-Petit value of $3R \approx 24.93 \ \mathrm{J}/(\mathrm{K}\cdot \mathrm{mol})$ adopted from thermodynamics of dilute gases. The specific heat of solid state matter reaches at melting temperatures the limiting value of $27-29\ \mathrm{J}/(\mathrm{K}\cdot \mathrm{mol})$, which is different for different materials. The behaviour of specific heat in the quasi-low temperature regime contain the universal quartic term, with the addition of a sample specific cubic contribution of the surface heat capacity.  However, the key ideas of P. Debye, the scaling expressed as a universal function of specific heat  and the velocity of sound as the main physical characteristic of heat, were correct. The field theory of specific heat \cite{Gusev-FTSH-RJMP2016} presents the mathematically correct implementation of these physical ideas.

False theories that have never been verified are presented in textbooks, despite the experimental progress with acquiring precision data of specific heat. We discussed in \cite{Gusev-FTSH-RJMP2016,Gusev-QLTBSH-RSOS2019} the misleading graphical proofs of the Debye theory presented in the very influential textbook of C. Kittel \cite{Kittel-book2005}. Let us address here another textbook, a monograph by M.T. Dove  on  the lattice dynamics \cite{Dove-book1993}. There, a comparison with experiment, based on the specific heat of sodium chloride, NaCl, is made with help of Fig. 5.5 on p. 76. The book claims that plotted experimental data (no publication reference is given) displays the cubic in temperature behaviour of the specific heat  at low temperatures, therefore, this graph is supposed to experimentally prove the cubic law of the Debye theory. 

We discuss here only the Debye theory because, the lattice dynamics of Born-von Karman does not predict any observable physical quantities, like the specific heat, it only fits its internal parameters (atomic interactions between nodes of a lattice) to the experimental data of acoustic frequencies \cite{Dove-book1993}, usually determined by the neutron scattering. Criticism of the theory of lattice dynamics was presented in theoretical \cite{Gusev-QLTBSH-RSOS2019,Raman-PIASA1941,Raman-PIASA1955} and experimental \cite{Brockhouse-RMP1958,Brockhouse-PRL1959} works. 

\begin{figure}[ht]
  \caption{Specific heat of sodium chloride, data of Ref. \cite{Morrison-PFS1956} as presented in Ref. \cite{Barron-PRSA1964}}
  \centering
\includegraphics[scale=0.7]{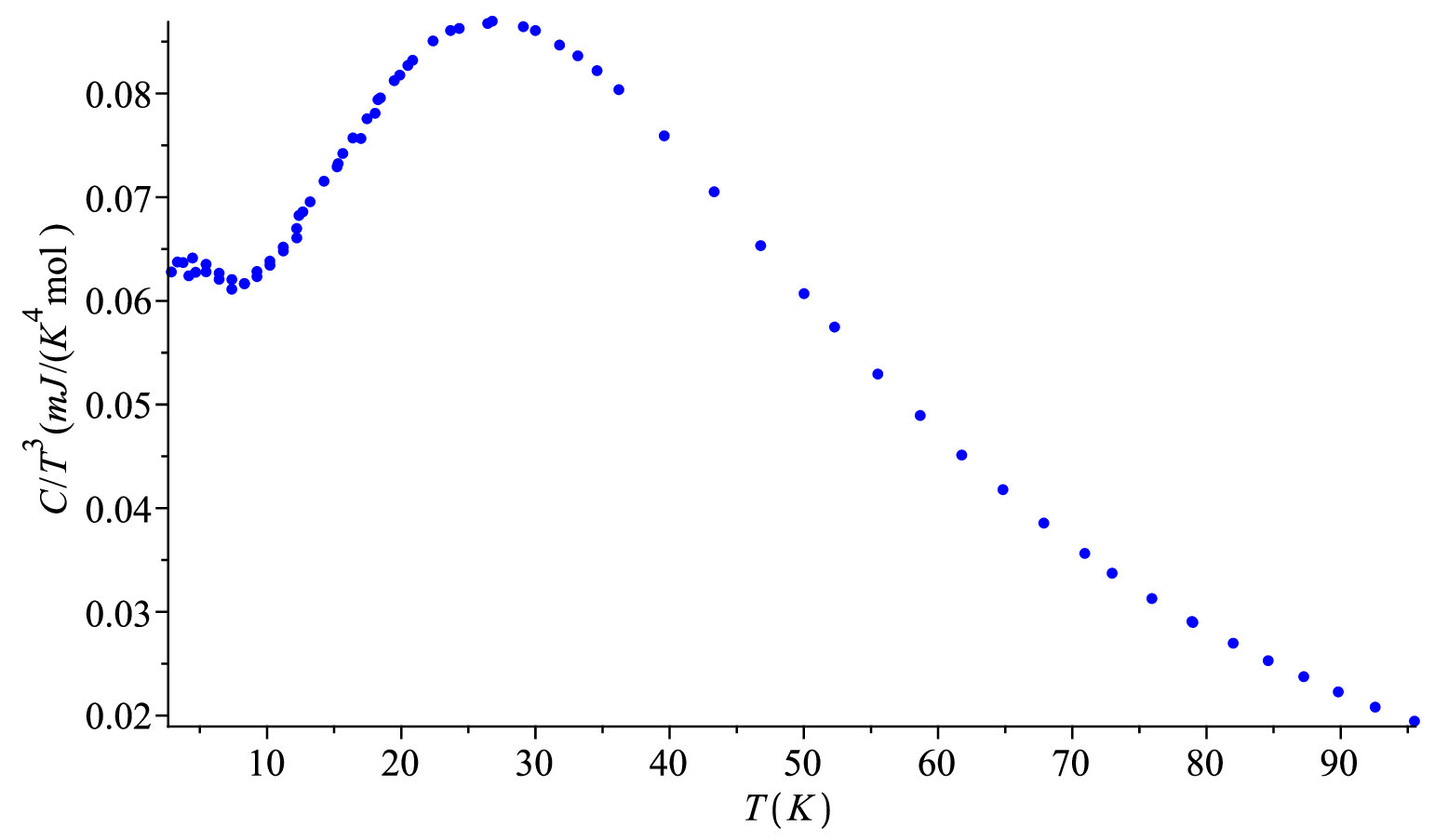}
\label{NaCl-Morrison-T3}
\end{figure}

We searched for publications on the specific heat of NaCl, some were cited and discussed in Sect. \ref{NaClpowder}, here we consider the  study  was done in the lab of J.A. Morrison \cite{Morrison-PFS1956}. Together with D. Patterson he measured the specific heat of sodium chloride for the range of temperatures from about 3 to 267 K. However, in that paper they presented only the analysis and the graph for these experiments. The actual experimental data were published much later in the appendix of Ref. \cite{Barron-PRSA1964}. These data represent a typical set of specific heat behaviour, as seen in the graph of $C/T^3$ vs $T$, Fig. \ref{NaCl-Morrison-T3}, where no cubic power law, which should be the {\em horizontal} straight line, is present, except at the very low temperature tail, which was already discussed.
In Fig. \ref{NaCl-Morrison-T3} the cubic law behaviour in the range of temperatures from 8 to 20 K, claimed in the textbook \cite{Dove-book1993}, is not observed. The approximately cubic behaviour at $T < 10\ K$ is explained in Sect. \ref{surface} as an evidence of the non-universal surface specific heat. 

The origin of this confusion in the literature on specific heat is clear: one may fit a data set with {\em any} fitting function and obtain some fitting coefficients, but one must always verify the statistical significance of the proposed function by calculating its statistic.  to select the best fitting function. Therefore, if raw data are treated, we can fit them with different power-law functions or their combinations, but different functions would have different statistical significance, as quantified by statistics like $\chi^2$ and standard errors.

The limit on the specific heat at higher temperature appears due to the cut-off imposed on the acoustic frequencies spectra of condensed matter by a finite inter-atomic distance \cite{Gusev-FTSH-RJMP2016}. Condensed matter  as ideal elastic media is a mathematical over-idealization, instead, matter has discrete, atomistic structure. This physical restriction limits the wavelengths of elastic waves in condensed matter. Properly incorporating this limit into a theory, as done in the field theory of specific heat, can resolve main troubles of the Debye theory.

The high temperature limit,  at the melting temperature, of the Debye theory is fixed to the value of $3R$, which is taken from the thermodynamics of gases. The vast experimental evidence demonstrate that the equipartition theorem is often not valid even in the heat theory of gases. Thus, it is not correct to consider it as the universal law and use it beyond its original scope of applications, dilute gases. The specific heat functions grow beyond this limit,  e.g. \cite{Passler-AIP2013,Gerlich-JAP1964}.

Anomalous 'deviations' from the Debye cubic law at low temperature, Chap. 2 of Ref. \cite{Dekker-book1960}, are not corrections to the Debye theory, they are indications of its  failure. The proof of this failure is the demonstration, with already available experimental data, of the fact that the cubic in temperature contribution of the specific heat is {\em not} universal, i.e. it is not a 'bulk' property of the given material, but it depends on a size and a shape of the body via its effective size. This was shown above with the specific heat of rock salt powders and natural diamonds. The dedicated precision measurements of  substances with various geometrical characteristics are urgently needed. Perhaps, reference data for specific heats should be measured again in view of this experimental evidence.

\subsection{Physical theory vs mathematical model}

The first two publications on the field theory of specific heat \cite{Gusev-FTSH-RJMP2016,Gusev-QLTBSH-RSOS2019} met the criticism of R. P\"assler, summarized in the dedicated paper \cite{Passler-PSSB2019}. It is encouraging to see the early interest in these ideas since one of the goals of our publications was to excite the critical revision of existing thermal theories by condensed matter physicists \cite{Gusev-FTSH-RJMP2016}, p. 71.  Despite its generally negative opinion,  the empirical analysis \cite{Passler-PSSB2019}  extends the exploration of experimental data we began in  \cite{Gusev-FTSH-RJMP2016,Gusev-QLTBSH-RSOS2019}, and its conclusions generally agree with the field theory of specific heat. 

The preliminary statistical analysis of \cite{Gusev-FTSH-RJMP2016} was improved in \cite{Gusev-QLTBSH-RSOS2019}. In fact, the work  \cite{Passler-PSSB2019} used the form of the statistical estimate, (\ref{linear}) and (\ref{quartic}) introduced in our second paper \cite{Gusev-QLTBSH-RSOS2019}. Nevertheless, we emphasize that the focus on an exact and universal power of temperature in experimental data at the QLT regime is misleading. We used this subject to highlight the differences from the textbook theories and to draw the attention  of the solid state physics community to the glaring problems in the condensed matter thermodynamics. For crystalline matter, which is anisotropic and support several longitudinal and transverse velocities of sound, the quartic law may be observed, among other terms, in a range of temperatures, we call the quasi-low temperature regime. The $T^4$ power is overcome by the cubic  law of surface heat (\ref{quartic}) below this temperature region and suppressed by the exponents of the universal thermal function (\ref{Theta}) above it.

The completed and calibrated field theory will include all geometrical (volume, surface, edge) contributions to specific heat and be derived with the full set of group velocities of sound that may exist in a condensed matter body. For example, the surface  specific heat, which is a $T^3$ function only in {\em its} QLT regime, but otherwise is a universal thermal function of surface heat (\ref{Theta2D}), qualitatively similar to the bulk one (\ref{Theta}), is  present at any temperature, but its {\em relative} contribution with respect to the bulk specific heat depends on temperature. The numerical analysis of measured specific heats in \cite{Passler-PSSB2019} supports this idea by finding 'sub-quartic' and 'super-quartic' behaviours. These behaviours may be determined by the full spectrum of acoustic frequencies of a crystal that was not incorporated yet into the field theory of specific heat. At the same time, the phenomena of surface heat capacity also must always be taken into account as shown in Sect. \ref{surface}, and it is obviously responsible for the polynomial (\ref{quartic}), which again may be more complex due to several transverse velocities.

Regretfully, the work \cite{Passler-PSSB2019} is devoted entirely to mathematical modelling, with the focus on the $T^3$ vs $T^4$ problem, and leaves theoretical physics out of consideration. The difference between the two is fundamental. A physical theory describes and {\em predicts} the functional behaviour of a physical quantity by a mathematical construction based on the input from observed physical quantities of {\em different kind}. For example, in the proposed geometrical formalism, the specific heat behaviour is derived from the sound velocities and the lattice constants, i.e. thermal properties are derived from mechanical ones. Instead the modelling is concerned with the best fit of {\em existing data} by mathematical functions that have no immediate relations to measured physical properties. 

The modelling of specific heat by extended polynomials performed in \cite{Gusev-QLTBSH-RSOS2019,Passler-PSSB2019}, was first introduced to the low temperature  physics by T.H.K. Barron and J.A. Morrison \cite{Barron-CJP1957} in order to describe experimental data obtained in their laboratory, because existed theories could not account for the observed behaviour of specific heat. However, the proposed fitting equation was just empirical, it was not justified by a physical theory. The diatomic linear chain model considered in the appendix of Ref. \cite{Barron-CJP1957}  cannot not serve as a theoretical foundation for the proposed specific heat function for the same reasons the lattice dynamics cannot be accepted as a physical theory, as discussed above and in \cite{Gusev-QLTBSH-RSOS2019}. Therefore, some  coefficients of such a polynomial are usually determined with unacceptably low statistical significance, i.e. the statistical hypothesis that the data correspond to the tested function is rejected. Unfortunately, sixty years ago physicists believed more in 'fundamental' theories and had less trust in statistical analysis.

Any scientific theory must not only describe existing but also predict unknown physical phenomena.  Thus, in \cite{Gusev-FTSH-RJMP2016} we made some  predictions for the specific heat of grey tin using only its known {\em mechanical} properties and {\em thermal} properties of  other materials in the same lattice class. These predictions are tested in Sect. \ref{tin}. Since the grey tin data were already published, but not known to us, this case could  be considered a blind calibration test. 

\subsection{Current state and future completion of the theory}

The physical meaning of the scaling is the mathematical nature (universality) of the fundamental object of the universal thermal sum (\ref{TF}). This is it, traditional thermodynamics could not be made independent of the material specific physical values. In our formalism, this dependence is reduced to the minimum, i.e. to a single value $T_0$. However, even this value could be avoided if one knew atomic and  elastic properties of a material.
In fact, the ultimate form of scaling could be observed in liquids because they are truly isotropic and (usually) possess only a pressure velocity of sound. This subject will be worked out next, and the completion of the theory of specific heat for crystalline matter is postponed, because it requires a better expertise in crystallography. 

The way to complete the field theory of specific heat, in our opinion, is to replace a scalar function of temperature by the tensor quantity. Indeed, as long as temperature got connected with a velocity of sound, it became a vector, because velocity is a vector. In anisotropic matter, the stress tensor defined by elastic moduli gives the tensor of velocities. The possible way to relate the new  tensor $\beta_{ij}$ with the temperature of traditional thermodynamics is to make some averaging, e.g. the determinant. Whether this method would be working could be seen through building specific models and calibrating them them with experimental data.

Over the last two decades extensive measurements the specific heat of glasses studied
their universal thermal behaviour, in particular, the convex shape of $C/T^3$ function, e.g. Fig. \ref{diamond-T3-size}, was dubbed as the 'boson peak' \cite{Nakamura-JPSJ2014}. The main idea of our work is that this universal scaling can be explained by the universal function.
The specific heat functions of glasses are qualitatively similar to crystals, i.e. their fundamental function is the same (\ref{Theta}). Experimental evidence forces us to discard the old belief that glasses have thermal properties different in principle from those of crystalline matter, e.g. \cite{Mertig-SSC1984}. The observed quantitative difference is explained by the isotropy of glasses and anisotropy of crystals. Consequently, glasses possess only two, longitudinal  and transverse, elastic waves, while crystals spectra are quasi-discrete, i.e. have several pronounced peaks, e.g. \cite{Huntington-book1958,Raman-PIASA1955}. These differences are further complicated by the surface heat capacity. Thus, it is quite natural to first complete the theory for amorphous matter, before proceeding to crystalline one. We hope that experts on the  physics of glasses will utilize the proposed mathematical ideas.

The approach opposite to thermodynamics shaped by field theory is statistical physics that operates with the notions and language of probability theory and statistics. However, statistical physics employed the combinatorics of indistinguishable particles that was not mathematically correct, which caused its discrepancy with experiment, especially in the area of real gases and phase transitions. Correct mathematics based on the number theory, required for a thermal theory of discrete matter, was created by  S. Ramanujan and G.H. Hardy \cite{Hardy-PLMS1918}   a century ago. Only relatively recently it has been implemented as a physical theory by V.P. Maslov \cite{Maslov-MN2013,Maslov-RJMP2015}. We expect that the field theory of specific heat can be worked out into a full theory of thermal phenomena that would not need statistics to be complete, in contrast to statistical thermodynamics. We also plan to show its consistency with the Maslov's statistical physics by describing, in a different way, the same physical phenomena. 

Mathematics of the evolution equation on Riemannian manifolds (spacetime), which is employed above, is universal (as is any mathematics). It was developed for almost a century, in parallel and independently, by mathematical physicists, e.g. \cite{Avramidi-book2015}, and by pure mathematicians, e.g. \cite{Ricci-flow-1-book2007}. This mathematics was supplemented with  a physical postulate that temperature can be introduced into a theory as the inverse of the Euclidean cyclic time. As any other physical postulate, it came from a historically long theoretical development based on experimental observations. The new, seemingly strange geometrical setup of the $\mathbb{R}^{d} \times \mathbb{S}^1$ spacetime, which is home for sound waves in condensed matter media, is a replacement for the traditional phase-space formalism, which is inhabited by moving material particles or immaterial quasi-particles. For generations physicists used to accept the six-dimensional space of coordinates and momenta and became comfortable with thinking within this abstract space. Let us hope a mathematical notion of the product of the real three-dimensional space with the imaginary cyclic time would also eventually become a common mathematical formalism.

Since physics aims to discover the most general properties of natural phenomena, its theories must embody the most general mathematics. There is nothing more general and fundamental in mathematics than the geometry of spacetime, which could be explored with the versatile tool of the evolution equation. The phenomenon of scaling, Sec. \ref{scaling}, discovered empirically and mathematically, is clearly one of the most fundamental properties of Nature. 
This discovery should be investigated and elaborated into practical models that would  aid engineers and technologists with analytical tools instead of data tables.

{\bf Conflict of Interest Statement}.
I have no conflict of interest at this time.

{\bf Data Availability Statement}.
Data sharing is not applicable to this article as no new data were created or analyzed in this study.

\section*{Acknowledgements}

I am  grateful to Prof. R.K. Kremer for sending me the data of the diamond's specific heat measured at Max Planck Institut f\"ur Festk\"orperforschung.


\end{document}